\documentclass[aps,preprintnumbers,superscriptaddress,twocolumn,amsmath,amssymb,floatfix,prb]{revtex4-1} 
\pdfoutput=1
\usepackage{graphicx,float}
\usepackage[usenames,dvipsnames]{color}
\usepackage[colorlinks,bookmarks=false,citecolor=NavyBlue,linkcolor=Red,urlcolor=blue]{hyperref}
\usepackage{enumitem} 
\usepackage[export]{adjustbox}

\usepackage{mathtools}

\newcommand\be{\begin{equation}}
\newcommand\bea{\begin{eqnarray}}
\newcommand\bes{\begin{subequations}}
\newcommand\esu{\end{subequations}}
\newcommand\ee{\end{equation}}
\newcommand\eea{\end{eqnarray}}

\newcommand{\cmmnt}[1]{}

\newcommand\ba         {\begin{eqnarray} } 
\newcommand\ea         {\end{eqnarray} } 

\usepackage{braket}

\newcommand\ocite[1]{[\onlinecite{#1}]}
\newcommand{\dd}{{\rm d}}

\usepackage{dsfont}

\begin{document}

\title{Universal late-time dynamics  in isolated one-dimensional statistical systems with topological excitations}

\author{Alvise Bastianello}
\affiliation{Institute for Theoretical Physics, University of Amsterdam, Science Park 904, 1098 XH Amsterdam, The Netherlands}
\author{Alessio Chiocchetta}
\affiliation{Institute for Theoretical Physics, University of Cologne, D-50937 Cologne, Germany}
\author{Leticia F. Cugliandolo}
\affiliation{Sorbonne Universit\'{e}, Laboratoire de Physique Th\'{e}orique et Hautes Energies, CNRS UMR 7589, 4, Place Jussieu, Tour 13, 5\`{e}me \'{e}tage, 75252 Paris Cedex 05, France}
\affiliation{Institut Universitaire de France, 1 rue Descartes, 75231 Paris Cedex 05, France}
\author{Andrea Gambassi}
\affiliation{SISSA - International School for Advanced Studies, via Bonomea 265, 34136 Trieste, Italy}
\affiliation{INFN, Sezione di Trieste, via Bonomea 265, 34136, Trieste, Italy}

\date{\today}


\begin{abstract}
We investigate the non-equilibrium dynamics of a class of isolated one-dimensional systems possessing two degenerate ground states, initialized in a low-energy symmetric phase.
We report the emergence of a time-scale separation between fast (radiation) and slow (kink or domain wall) degrees of freedom. We find a universal long-time dynamics, largely independent of the microscopic details of the system, in which the kinks control the relaxation of relevant observables and correlations.
The resulting late-time dynamics can be described by a set of phenomenological equations, which yield results in excellent agreement with the numerical tests.
\end{abstract}

\maketitle

\section{Introduction}
\label{sec_intro}

Understanding the non-equilibrium dynamics of quantum and classical many-body systems represents a formidable challenge. In contrast to thermal equilibrium, for which thermodynamics and statistical mechanics provide efficient computational tools, no systematic description exists for non-equilibrium systems. As a consequence, while simplifications may exist in specific cases, one is in principle forced to solve the equations of motion of a macroscopic number of degrees of freedom.

In this respect, instances in which simplified and universal non-equilibrium scale-invariant behaviors emerge are particularly valuable, as in the case of systems undergoing critical relaxation or coarsening in contact with a bath\cite{Hohenberg1977,Tauber_book,Henkel_book}, 
which were thoroughly investigated in the past years. 
More recently, there has been a rising interest in exploring analogous phenomena in the dynamics of isolated quantum systems\cite{Greiner2002,Hackermller1621,Trotzky2012,Cheneau2012,Langen2013,PhysRevLett.111.053003,Bloch2005,Joerdens2008,Kinoshita2006,Hofferberth2007,Gring1318,PhysRevLett.110.205301,PhysRevA.91.043617,PhysRevLett.115.085301,PhysRevLett.115.175301,PhysRevLett.103.150601,Paredes2004,Fukuhara2013}. 
In part,  this was motivated by the impressive progress in experiments with ultracold atoms that made possible the realization of almost perfectly isolated quantum many-body systems, and the monitoring of observables in real time.

The dynamics of isolated statistical systems, either classical or quantum, is  less well understood and, in a sense, more challenging than that of systems in contact with reservoirs. 
On the quantum side, the dynamics eventually leading to the thermalization of isolated systems is a very active area of research. Our current understanding of whether 
they thermalize or not is primarily based on the eigenstate thermalization hypothesis (ETH) 
\cite{PhysRevA.43.2046,PhysRevE.50.888,Rigol2008,PhysRevLett.103.100403,PhysRevLett.108.110601,Reimann_2015,Deutsch_2018} which roughly states that almost all the eigenstates behave as if they were thermal. 
Important exceptions to ETH exist, such as integrable models \cite{Calabrese_2016} and the recently observed quantum scars \cite{Turner2018,PhysRevB.99.161101,PhysRevLett.122.040603,PhysRevLett.122.173401,PhysRevB.98.155134,PhysRevLett.122.220603}, or strongly disordered systems which display many-body localization \cite{MBL_rev}.

Besides the steady state itself, understanding the late-time dynamics and the eventual  approach to a thermal ensemble, in those cases in 
which this happens, is an even more compelling challenge. 
Within the realm of interacting (and not integrable\cite{Calabrese_2016}) systems, the intimately non-perturbative nature of the process makes extremely hard to achieve an analytic description. 
Some progresses in this direction
can be made in weakly-interacting models through Boltzmann-like equations
\cite{PhysRevLett.100.175702,PhysRevLett.115.180601,PhysRevB.94.245117,PhysRevE.86.031122,PhysRevE.88.012108,PhysRevB.95.104304}. 
Beyond these, recent exact results have been obtained in specific strongly interacting models, known as quantum circuits \cite{PhysRevX.7.031016,PhysRevX.8.021014,PhysRevX.8.021013,PhysRevX.8.041019,PhysRevLett.121.060601,PhysRevX.8.031058,PhysRevX.8.031057,PhysRevLett.121.264101,PhysRevX.9.021033,2019arXiv190402140B}.

Scaling phenomena and universality have been instrumental in the description of dissipative macroscopic systems in and out of equilibrium, especially in connection with collective or coarsening phenomena \cite{Hohenberg1977,Tauber_book,Henkel_book}. 
One could then naturally expect that they might play a crucial role also in the dynamics of isolated quantum systems. Indeed, the possible emergence of a scaling behavior should dramatically simplify their description, as the relevant aspects of the long-time, large-distance features of a system would then become insensitive to its microscopic details. 
In addition, non-equilibrium collective dynamics typically results in a divergence of relaxational time scales, thus delaying or even hindering thermalization\cite{Bray1994,Gambassi2005,Henkel_book}.

In isolated systems, a variety of novel universal dynamics have been theoretically predicted
\cite{Langen2016} and experimentally observed \cite{Nicklas2015,Pruefer2018,Erne2018}. In particular, a simple protocol which results into a dynamical universal behavior consists in quenching a system across a critical point starting from a disordered phase. The subsequent \emph{coarsening} dynamics, characterized by the formation of spatial domains of different ordered phases and diverging relaxation times \cite{Bray1994,Cugliandolo2015}, has been shown to occur also for isolated Hamiltonian 
dynamics\cite{Damle1996,Mukerjee2007,Lamacraft2007,Williamson2016a,Williamson2016b} and in certain cases it is accompanied by a novel behavior, with no counterpart in the presence of thermal 
baths\cite{Sondhi2013,Sciolla2013,Maraga2015,Staniscia2019}. 
Glassy features with similarities and differences from the ones found under dissipative 
dynamics\cite{Cugliandolo2002} have also been exhibited in solvable 
models\cite{Nessi1,Nessi2,Nessi3}. 

This scenario is strongly affected by the spatial dimensionality $d$ of the system, as it can be understood on the basis of simple thermodynamic arguments. In fact, in the presence of short-range interactions (assumed henceforth), the growth of a domain of one phase within another one costs an energy which is proportional to the area of the interface.  While in $d>1$ the increase of the domain size also requires a growth of the extension of interface, this is not the case in one spatial dimension $d=1$: regions within a different phase can grow up to thermodynamic scales, paying only a finite energy price.
Indeed, one-dimensional systems with short-range interactions cannot sustain finite-temperature phase transitions \cite{huang2009introduction}. 
However, recent experiments \cite{Erne2018,Pruefer2018} showed the emergence of non-equilibrium universal dynamics in isolated quasi one-dimensional Bose gases, demonstrating that these kind of systems may still display rich and largely unexplored phenomena.

In this work, we investigate the non-equilibrium dynamics of isolated one-dimensional systems, reporting the natural emergence of a time-scale separation between fast and slow degrees of freedom --- the latter being related to topological excitations (i.e., \emph{kinks} or domain walls) --- which, in turn, results into a universal long-time, large-distance dynamics, largely independent of the microscopic details of the system.
In particular, we consider a system with an order parameter $\phi$ and a potential with $\mathbb{Z}_2$ symmetry. With $\pm\bar{\phi}$ we denote the two minima of that potential, i.e., the two zero-temperature phases, which we refer to as vacua.
We assume that the system is initially prepared at time $t=0$ in a low-energy state with $\mathbb{Z}_2$ unbroken symmetry (i.e., $\langle\phi\rangle=0$) and short correlation length. 

Then, the system is let evolve in isolation for $t>0$, with a Hamiltonian which admits two phases at zero temperature, connected by the $\mathbb{Z}_2$ symmetry.
Such a non-equilibrium protocol is usually referred to as \emph{quench}, which has recently attracted a lot of attention in the context of quantum statistical systems\cite{Calabrese_2007}.
At low energy, topological excitations interpolating between these two phases, known as kinks, play a central role.
In equilibrium, kinks are known to lead to universal features\cite{kinkbook}, for example the correlation function of the order parameter is entirely determined by the mean kink density, independently of the microscopic structure of the model\cite{PhysRevB.11.3535}.
Out of equilibrium, this feature partly carries over to integrable models \cite{PhysRevE.93.062101,PhysRevLett.119.100603,PhysRevB.100.035108,PhysRevLett.119.010601}. A small, symmetry-breaking term in the Hamiltonian which controls the dynamics of the system may cause the confinement of the topological excitations, which has been understood to have strict connections with the quantum scars\cite{Kormos2016,PhysRevLett.122.130603,PhysRevB.99.195108,PhysRevB.99.180302}. 

Here, on the contrary, we focus on the case without explicit symmetry-breaking or confinement. 
After its initial non-equilibrium evolution, the system is eventually expected to thermalize at a low, but non-zero temperature, determined by the energy initially injected into the system. The main results of our analysis can be summarized as follows: 
\begin{enumerate}[label=(\roman*)]
\item The large-distance behavior of the correlation functions of the order parameter is still governed by the mean kink density also when the system is out of equilibrium. 
The associated correlation length is proportional to the inverse kink density which, at low energies, is much larger than any other length scale in the system and thus controls the 
large-distance behavior.
\item  After an initial transient, we observe that the time scales of relaxation of the kinks become significantly larger than those of other excitations. The total density of kinks, in fact, relaxes on a time scale which becomes exponentially large upon increasing the final inverse temperature, thus determining the entire long-time dynamics. 

\item The emergence at long times of slow and collective degrees of freedom is revealed by the dynamics of observables which are even under the $\mathbb{Z}_2$ symmetry, which in practice are unable to distinguish between the two phases and are therefore insensitive to topological excitations.
Nevertheless, we observe that, at late time, the difference between the time-dependent value of the observable and its final one in thermal equilibrium is proportional to the difference between the instantaneous kink density and its final equilibrium value.
This relationship does not depend on the details of the initial state and the involved proportionality constant is completely determined by the thermal equilibrium ensemble.
\item The dynamics of the kink density is found to be determined by a simple phenomenological equation, involving two parameters: the equilibrium kink density corresponding to the final temperature and a cross-section which describes the probability that two kinks are annihilated in a scattering event. 
This cross-section turns out to be solely fixed by equilibrium properties.
\end{enumerate}

Kinks are intrinsically non-perturbative excitations: since the corresponding field interpolates between the two vacua, the kink configurations are not perturbatively close to any of them, making analytical calculations extremely hard. 
This is the reason why state of the art quantum field-theoretical approximations (such as the 
2PI formalism\cite{Berges2004rev}) have been shown to fail in capturing the effects of topological defects\cite{Rajantie2006,Berges2011}.

In order to circumvent these difficulties, we focus here on the classical world as a convenient arena for our investigation: large scale ab-initio simulations of the microscopic model are easily performed, backing up our phenomenological reasoning. 

Accordingly, we mostly present and discuss our results referring to classical systems and only briefly comment on the quantum case, but we provide arguments supporting the fact that the general mechanism causing the mentioned universal behavior.

Furthermore, while the proposed non-equilibrium protocol can be implemented and investigated in purely classical models, under the proper conditions which will be discussed later on, classical systems can be viewed as a good approximation of the quantum theory. This observation further supports the generality of the proposed picture with respect to the actual classical or quantum nature of the statistical system.

The paper is organized as follows. Section~\ref{sec_themodel} introduces the model of interest, providing details on the spectrum of excitations (Sec.~\ref{sec_zeroT}) and on its low-temperature equilibrium thermodynamics (Sec.~\ref{sec_thermo}).
Section~\ref{sec_OUT} addresses the non-equilibrium behavior of the system: based on the insight in the 
low-temperature physics built in the previous Section, we discuss the non-equilibrium behavior and the emergence of the time-scale separation, which are then thoroughly checked via numerical calculations.
In Sec.~\ref{sec_concl} we present our conclusions, while the numerical methods used in our investigation are reported in two short Appendices.

\section{The model: low energy excitations and thermodynamics}
\label{sec_themodel}

Among the wide class of classical or quantum models supporting topological excitations, we consider 
a chain of anharmonic oscillators governed by the Hamiltonian
\be\label{eq_H}
\!\!
H= a \! \sum_x 
\left[\frac{\Pi^2(x)}{2}+\frac{(\phi(x+a)-\phi(x))^2}{2a^2}+V(\phi(x))\right],
\ee
where $\phi(x)$ indicates the displacement of the oscillator at position $x$ along the chain, $\Pi(x)$ is the momentum conjugated to $\phi(x)$, while $a$ stands for the lattice spacing and the sum runs over the sites of the chain. 
The potential $V(\phi)$ is assumed to be $\mathbb{Z}_2$-symmetric, i.e.,  $V(\phi)=V(-\phi)$ and shaped as a double well with two minima at the two vacua $\pm \bar{\phi}$. 
A prototypical form of $V$ is given by $V(\phi)=\lambda (\phi^2-\bar{\phi}^2)^2$, but the results discussed further below apply to any shape of $V$ with the same features.
Without loss of generality, we assume $V(\bar{\phi})=0$, possibly by adding a constant offset to the potential.

The Hamiltonian~\eqref{eq_H} can be regarded either as a classical function or a quantum operator, 
by imposing either canonical Poisson brackets on the conjugate fields $\{\phi(x'),\Pi(x)\}=a^{-1}\delta_{x,x'}$ or 
canonical commutation relations $[\hat \phi(x'),\hat \Pi(x)]=ia^{-1}\delta_{x,x'}$ (with the convention $\hbar=1$ and the fields upgraded to operators), respectively. 
In Eq.~\eqref{eq_H} it might be convenient to take the limit  $a\to 0$ of vanishing lattice spacing: by approximating the hopping term with a derivative, one reaches the continuum limit with Hamiltonian
\be\label{eq_H_continuum}
H=\int \dd x\, \left[ \frac{1}{2} \Pi^2(x)+\frac{1}{2}(\partial_x \phi)^2+V(\phi(x))\, \right].
\ee
For the sake of simplicity, in the following we discuss the thermodynamics of the model mostly in the the continuum and classical case, as in Eq.~\eqref{eq_H_continuum}, discussing the effects of a possible quantization or of the underlying lattice in Secs.~\ref{sec_quantum_effects} and \ref{sec_lattice_effects}, respectively. The numerical analysis will be performed on the classical lattice model in Eq.~\eqref{eq_H}.

\subsection{Low-energy excitations}
\label{sec_zeroT}

Under the assumptions discussed in the previous Section, the classical Hamiltonian $H$ in Eq.~\eqref{eq_H} admits two degenerate minima (or vacua) and correspond to $H=0$. A field configuration initialized in one of the two vacua $\phi(x) = \pm \bar \phi$ does not evolve in time.
At low energies, instead, $H$ admits two different species of excitations: (a) topological excitations, i.e., \emph{kinks}, which interpolate between the two vacua \cite{kinkbook}, and
(b) field fluctuations occurring around the same phase, which we refer to as \emph{radiation} \cite{kinkbook}. 
Below we shortly describe these two kind of excitations before considering their effects on the low-temperature 
thermodynamics of the model.

Topological excitations naturally emerge as finite-energy solutions of the equation of motion %
\be\label{eq_kinkeq}
\partial_t^2\phi-\partial_x^2\phi+V'(\phi)=0,
\ee
associated with the continuum Hamiltonian in Eq.~\eqref{eq_H_continuum}.
Those with the lowest (kinetic) energy do not evolve in time, have $\partial_t\phi=0$ and therefore they are described by the first-order differential equation
\be\label{eq_kinkprof}
(\partial_x\phi)^2- 2V(\phi)=0,
\ee
which is derived after simple manipulations.
In addition to the trivial solutions $\phi(x)=\pm \bar{\phi}$, which pin the field value to one of the two minima of  $V$, there is always a non-trivial solution $\phi^K(x)$ interpolating between the $-\bar{\phi}$ at $x\to-\infty$ and $\bar{\phi}$ at $x\to +\infty$ (kink) or the other way (antikink).
Note that the kink is non-local with respect to $\mathbb{Z}_2$-odd observables $\mathcal{O}_{\text{odd}}(\phi)$ such that $\mathcal{O}_{\text{odd}}(-\phi) = - \mathcal{O}_{\text{odd}}(\phi)$ because $\lim_{x\to -\infty}\mathcal{O}_{\text{odd}}(\phi^K(x))\ne \lim_{x\to +\infty}\mathcal{O}_{\text{odd}}(\phi^K(x))$. On the other hand, it is local when $\mathbb{Z}_2$-even observables are considered. 
For example, the energy density is localized around the center of the kink $x_0$, defined as $\phi^K(x_0)=0$. Without loss of generality, one can choose $x_0=0$.
In this case, the antikink profile  $\phi^{AK}(x)$ is related to that of the kink $\phi^{K}(x)$ by a spatial reflection with respect to the origin (or, more generally, to $x_0$): $\phi^{AK}(x)=\phi^{K}(-x)$.

Due to the translational invariance of the Hamiltonian $H$, any kink solution $\phi^K(x)$ can be arbitrary translated along the real line, without affecting the corresponding value of the energy. 
In addition, a Lorentz boost of the space-time coordinates $(t,x)$ allows one to write the solution of Eq.~\eqref{eq_kinkeq} corresponding to a moving kink, starting from a static one. 
In this respect, one can interpret the resulting kink 
\be
\phi_{v,y}^K(t,x)=\phi^K(\gamma(x-y-vt))\, ,
\ee
as a particle-like excitation of the chain with a well-defined position $x_0$ and moving with a certain velocity $v$,  where $\gamma = \gamma(v) =1/\sqrt{1-v^2}$. 
This picture is further confirmed by evaluating the energy $H[\phi_{v,y}^K]$ associated with $\phi_{v,y}^K$, 
which turns out to be that of a relativistic particle of mass $M$, i.e., $H[\phi_{v,y}^K]=\gamma M$, where $M$ is the $v$-independent quantity
\be\label{eq_rest_mass}
M \equiv \int_{-\bar{\phi}}^{\bar{\phi}}\!\dd \phi\, \sqrt{2 V(\phi)}.
\ee
In order to calculate $M$ according to this definition, one does not actually need the know of the kink profile, but only the shape of $V(\phi)$.
For the choice
$V(\phi)=\lambda (\phi^2-\bar{\phi}^2)^2$, 
the kink profile has a simple analytical expression\cite{kinkbook}
\be\label{eq_kink_phi4}
\phi^K(x)=\bar{\phi}\tanh\left(x \bar{\phi} \sqrt{2\lambda}\right),
\ee
and the mass obtained from Eq.~\eqref{eq_rest_mass} is $M=4\sqrt{2\lambda}\bar{\phi}^3/3$. 
Note that, generically, it is natural to define a kink width $w$ as the typical length scale over which the transition between the asymptotic values $\pm\bar\phi$ occurs: from the explicit expression in Eq.~\eqref{eq_kink_phi4}, for example, one can identify $w$ as $w = 2/(\bar\phi\sqrt{2\lambda})$. 

A single-kink solution is the lighter excitation interpolating between the two different vacua, 
therefore it is stable under time evolution\cite{kinkbook}, i.e., it cannot decay in lighter (less energetic) excitations: this fact will be essential in the late time non-equilibrium dynamics, as we will see later on.
In general, field configurations containing more than one kink or antikink can also be realized, but they are no longer stable under time-evolution.
For example, initial multi-kink configurations can be constructed by alternatively placing kink and antikink profiles along the real line. For periodic boundary conditions, the number of kinks and antikinks must be equal and we assume they are initially well separated from each other and with velocities and positions $(v_i,y_i)$, $i=1, \ldots, 2N$.
Such a multi-kink configuration is therefore given by
\be\label{eq_mK}
\begin{split}
& \Phi_{\{v_i,y_i\}_{i=1}^{2N}}(t,x)=\\
& 
\qquad\quad
\bar{\phi}\prod_{i=1}^N \left[\frac{\phi_{v_{2i-1},y_{2i-1}}^K(t,x)}{\bar{\phi}} 
\frac{\phi^{AK}_{v_{2i},y_{2i}}(t,x)}{\bar{\phi}}\right],
\end{split}
\ee
Since a kink must be followed by an antikink, at $t=0$ we require $y_{i+1} - y_i \gg w$, with $w$ the kink's width. 
Under this assumption of ``dilution" of kinks, each of them contributes additively to the total energy $H[\Phi_{\{v_i,y_i\}_{i=1}^{2N}}]=\sum_{i=1}^{2N}\gamma(v_i) M$ plus small corrections due to a short-range force between the kinks, which turns out to be attractive\cite{kinkbook}.

\begin{figure}[t!]
\includegraphics[width=0.8\columnwidth]{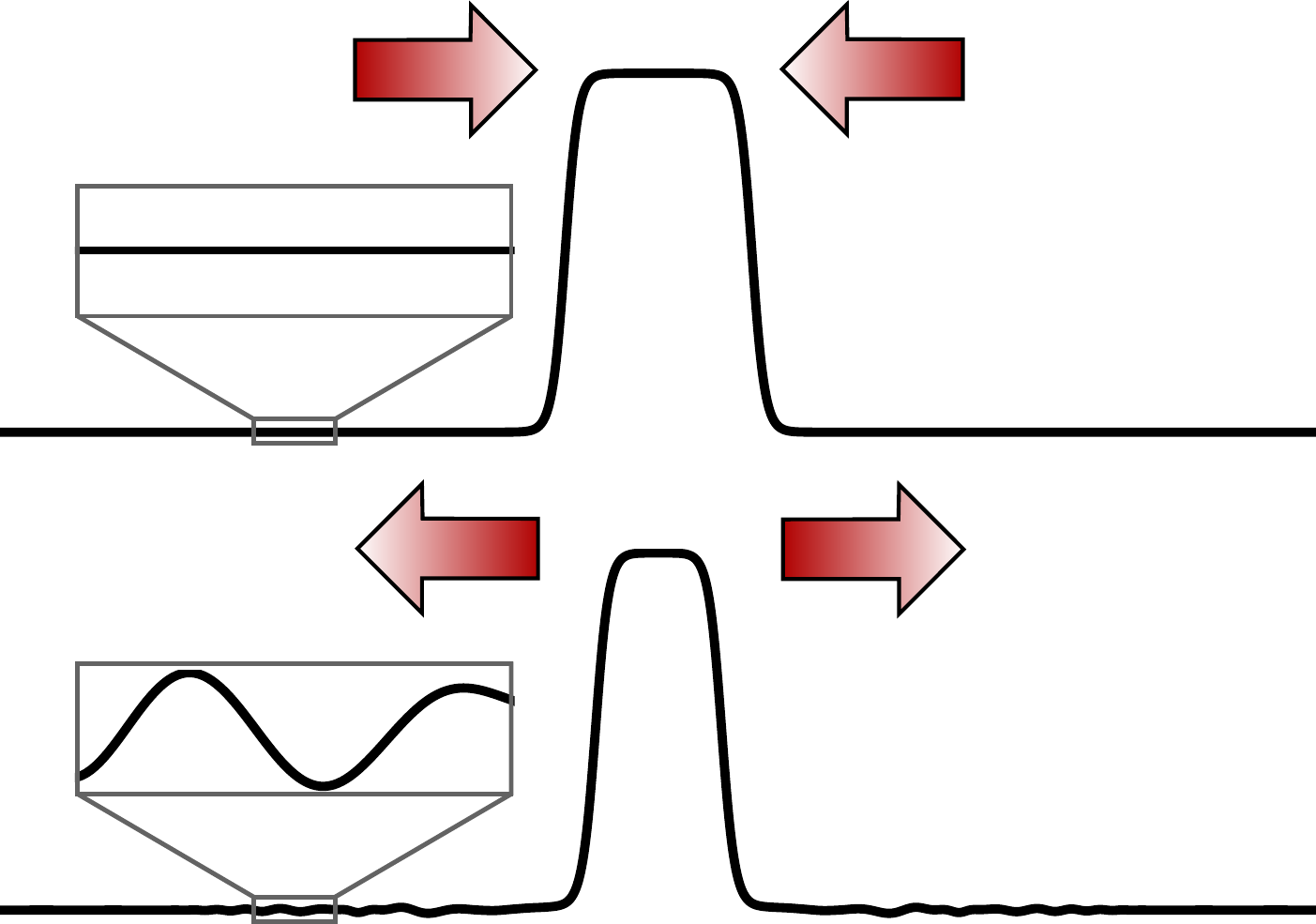}
\caption{\label{fig_1} (Color online) Cartoon of a kink-antikink approach (top) and collision (bottom), which results into the production of radiation after scattering has occurred.}
\end{figure}

In addition to kinks and antikinks, one can also construct excitations 
which consist of local fluctuations of the value of the field $\phi$ around one of the two minima of the potential. 
These excitations are usually referred as \emph{radiation}\cite{kinkbook}. 
By assuming such fluctuations to be small, the equation of motion \eqref{eq_kinkeq} can be linearized by expanding it around $\bar{\phi}$:
\be\label{eq_rad}
\partial_t^2\phi-\partial_x^2\phi+V''(\bar{\phi})(\phi-\bar{\phi})+\mathcal{O}((\phi-\bar{\phi})^2)=0.
\ee
This equation admits a mode decomposition in terms of plane waves characterized by a wave-vector $k$ (momentum) and a corresponding ``relativistic'' energy $\epsilon(k)=\sqrt{k^2+m^2}$, where the ``mass'' $m$ of these excitations is set by the curvature of the potential according to $m^2=V''(\bar{\phi})$.
While we introduced above kinks and radiation separately, these excitations are non-trivially coupled by the dynamics.
For example, if one lets the configuration~\eqref{eq_mK} evolve, the kinks and antikinks will move ballistically as free particles with the corresponding velocities. 
Generically, at some instant of time a kink profile will overlap with a neighbouring antikink one, undergoing scattering.
Given that a kink/antikink pair is no longer a stable solution of the equation of motion,
it is expected to couple with the sector of radiative excitations: part of the energy stored in the kink/antikink pair when they are initially well-separated is transferred to the radiation upon scattering, as depicted in Fig.~\ref{fig_1}. 
%
%
%
%

In general, there are no symmetries which prevent the coupling between the sector of the spectrum with two kinks and that one with radiation: accordingly, transitions between the two may happen.
In particular, a scattering event in which a kink-antikink pair annihilates and its energy is completely converted into radiation is possible. 
Because of the invariance of the equation of motion of the system under time-reversal, 
the reversed process is also possible. 
An important exception to this generic scenario is provided by integrable models \cite{claro2012nonlinear,PhysRevLett.56.2233,De_Luca_2016} in which infinitely many conservation laws can guarantee the stability of multi-kink configurations.
It can be shown that the only integrable model having the form of Eq.~\eqref{eq_H_continuum} and possessing topological excitations is the sine-Gordon model \cite{10.1007/BFb0105279} with $V(\phi)=(m^2/g^2)[1-\cos(g\phi)]$. 
In this work, we will not consider such a special case. In fact, the coupling existing between multi-kink configurations and radiation and the associated possibility of converting the first into the latter and vice-versa, 
are crucial in determining the non-equilibrium features of the model discussed below.

\subsection{Low-energy thermodynamics: decoupling kinks and radiation}
\label{sec_thermo}

In this Section we analyse the low-temperature statistical properties of the model in Eq.~\eqref{eq_H_continuum}, encoded in the classical partition function 
$\mathcal{Z}[\beta] = \int [\dd \phi] e^{-\beta H[\phi]}$ (with the corresponding expression in the quantum case) for $\beta \to\infty$.
An extensive literature has been dedicated to this 
problem \cite{PhysRevB.11.3535,PhysRevB.22.477,SahniMaz79} in the effort of obtaining a rigorous description. While we refer the reader to these contributions for technical details, here we only recall the basic ideas behind those studies.
A given field configuration $\phi(t,x)$ is split into a multi-kink solution similarly to Eq.~\eqref{eq_mK}, plus fluctuations $\chi(t,x)$ assumed to be small compared to the kink amplitude $2\bar\phi$, i.e.,
\be \label{rad_spl}
\phi(t,x)=\Phi_{\{v_i,y_i\}_{i=1}^{2N}}(t,x)+\chi(t,x).
\ee
If one imposes periodic boundary conditions at the boundaries of the large but finite chain, the system must contain an equal number of kinks and antikinks. However, this constraint is negligible in the thermodynamic limit.
Using the decomposition of $\phi$ as in Eq.~\eqref{rad_spl}, the corresponding energy $H$ (see Eq.~\eqref{eq_H_continuum}) naturally splits into three terms $H=H_K+H_\chi+U$, where $H_K$ is the energy of the independent kinks $H_K=\sum_{i=1}^{2N}\gamma(v_i) M$, $H_\chi$ is the Hamiltonian associated with small fluctuations around the vacua $\pm\bar\phi$, i.e., $H_\chi=\int \dd x \, [(\partial_t\chi)^2+(\partial_x\chi)^2+m^2\chi^2]/2$, and $U$ contains all the remaining interactions. 
In particular, $U$ accounts mostly for three kinds of interactions: those involving solely kinks, those of kinks with radiative modes, and finally the self-interactions of the radiation, which are determined by anharmonic corrections to Eq.~\eqref{eq_rad}.
At very low temperatures, the interaction $U$ can be neglected in a first, crude approximation\cite{PhysRevB.11.3535} and therefore the partition function $\mathcal{Z}[\beta]$ approximately factorizes as
\be\label{eq_Z_split}
\mathcal{Z}[\beta]\simeq \mathcal{Z}_K[\beta]\mathcal{Z}_\chi[\beta],
\ee
where $\mathcal{Z}_\chi[\beta]$ is the partition function for the quadratic Hamiltonian $H_\chi$ and $\mathcal{Z}_K[\beta]$ is the partition function of the free kinks with Hamiltonian $H_K$, considered as a collection of free relativistic impenetrable particles with mass $M$, size $w$, momenta $p_i=M\sinh\theta_i$, and energy $E=M\cosh\theta_i$, where $\theta_i$ is the relativistic rapidity associated with the velocity $v_i$ via $v_i = \tanh \theta_i$.
In particular, the average thermal kink density $n_\beta$  is readily computed
in the thermodynamic limit and for $\beta \gg M^{-1}$,
\be\label{eq_n_beta}
n_\beta\simeq w^{-1}\int_{-\infty}^\infty \dd \theta\,  e^{-\beta M \cosh\theta}\simeq w^{-1}\sqrt{\frac{2\pi}{\beta M}} e^{-\beta M}. 
\ee
Accordingly, the kink density is exponentially small at small temperatures.
In contrast, the fluctuations $\langle \chi^2\rangle$ of the radiation are suppressed only algebraically upon decreasing $\beta^{-1}$ towards zero. 
A crude estimate of the two-point spatial correlation function of the radiation field $\chi$ can be obtained by computing it on the thermal state based on the corresponding free Hamiltonian $H_\chi$, i.e.,
\be\label{eq_chi_corr}
\langle\chi(x)\chi(y) \rangle\simeq \int_{-\infty}^{+\infty} \!\!\frac{\dd k}{2\pi} \frac{e^{ik(x-y)}}{\beta\epsilon^2(k)}=\frac{e^{-m|x-y|}}{2m \beta},
\ee
where $\epsilon(k)$ was given after Eq.~\eqref{eq_rad}.
The attempts to verify numerically Eq.~\eqref{eq_n_beta} via Monte Carlo methods has a long history (see, e.g., Refs.~\ocite{KOEHLER19751515,PhysRevE.48.4284,GRIGORIEV198867,PhysRevLett.63.2337,PhysRevLett.68.1645} for the earliest works), revealing a series of difficulties.
First, large system sizes are needed in order to have enough statistics on the kinks since, as already mentioned, their density is exponentially suppressed upon decreasing the temperature.
Second, at finite temperature, the interaction cannot be neglected and the validity of Eq.~\eqref{eq_Z_split} becomes questionable. 
For instance, the radiation renormalizes the (bare) mass 
$M$ of the single kink\cite{PhysRevB.22.477} at the lowest order in the interaction strength. In turn, this renormalization affects significantly Eq.~\eqref{eq_n_beta}, in view of the exponential dependence of $n_\beta$ on such a mass.
Third, the number of kinks present in a field configuration at finite temperature is difficult to be determined in practice because, for example, large field fluctuations due to radiation might be misinterpreted as tight kink-antikink pairs. 

In view of the difficulties in finding a non-ambiguous definition of the kink density, it is important to identify  observables which are sensitive to the presence of topological excitations. Ideal candidates are correlation functions of observables  which are odd under the $\mathbb{Z}_2$ symmetry, for example the order parameter $\phi(x)$ itself.
In this case, consider $\langle \phi(x)\phi(y)\rangle$: at low temperatures its large-distance behavior turns out to be completely determined by the kink density $n_\beta$\cite{PhysRevB.11.3535}. In fact, one finds
\be\label{eq_twopt_thcor}
\langle \phi(x)\phi(y)\rangle =A_\beta e^{-2n_\beta |x-y|},
\ee
where $A_\beta>0$ is a temperature-dependent constant.
Although the derivation of this expression in thermal equilibrium is a textbook exercise \cite{PhysRevB.11.3535}, its generalization to non-equilibrium conditions plays a central role in our investigation and therefore we pause here to derive Eq.~\eqref{eq_twopt_thcor}, discussing the hypothesis under which this can be done.
As a starting point, we neglect the radiation and consider two points $x$ and $y$ (with, say, $x>y$) at a distance $x-y$ larger than the width of the kinks.
Let us then consider the observable $\phi(x)\phi(y)$ evaluated on a certain multi-kink configuration. Since each kink is responsible for a ``jump'' of the field from $-\bar{\phi}$ to $\bar{\phi}$ (viceversa for an antikink), then $\phi(x)\phi(y)$ approximately takes either the value $\bar{\phi}^2$ or $-\bar{\phi}^2$ depending on whether an even or an odd number of kinks is present in between the two points. 
At low temperatures, kinks are uniformly and independently distributed in space with a given density $n_\beta$. In this limit, we ignore the correlation existing among kinks and therefor all the terms of order $\sim n_\beta^2$. Under these assumptions, the probability $p_{[x,y]}(\ell)$ of having $\ell$ kinks within the interval $[x,y]$ (assuming $x>y$) is a simple Poisson distribution
\be
p_{[x,y]}(\ell)=\frac{(n_\beta|x-y|)^\ell}{\ell!}e^{- n_\beta|x-y|}\, .
\ee
Then, the two-point correlator of the order parameter can be easily computed as
\be\label{eq_corr_ord}
\langle \phi(x)\phi(y)\rangle\simeq\bar{\phi}^2\sum_{\ell=0}^\infty (-1)^\ell p_{[x,y]}(\ell)=\bar{\phi}^2e^{- 2n_\beta|x-y|}.
\ee
In order to estimate the effects of the radiation on the previous expression, it is useful to consider the correlation function of the radiation in Eq.~\eqref{eq_chi_corr} at low temperatures. 
From its functional form, it follows that the radiation develops and induces correlations up to a typical length scale $\sim m^{-1}$. Accordingly, for $n_\beta\ll m$ the kinks dominate the long-distance behavior of the correlation function of the order parameter.
In addition, the presence of the radiation affects the assumptions behind Eq.~\eqref{eq_corr_ord}, i.e., that
(i) the kinks are uniformly distributed and uncorrelated and 
(ii) the topological excitations make the field jump between $\pm \bar{\phi}$.
If $n_\beta\ll m$, the average distance between consecutive kinks is typically much larger than the correlation length of the radiation which, accordingly, cannot correlate them and they remain independently distributed. However, the radiation generically modify the single kink in such a way that the 
effective plateau value reached for $x\to \pm\infty$, approximately equal to $\pm\bar\phi$ changes. This renormalization affects the proportionality constant in Eq.~\eqref{eq_corr_ord}, but not the exponential decay.

Equation~\eqref{eq_twopt_thcor} also provides a useful indirect way to test the validity 
of Eq.~\eqref{eq_n_beta} and to study the finite-temperature corrections.
In fact, expectation values of observables on classical thermal states can be efficiently computed via the transfer matrix approach\cite{PhysRevB.6.3409} (briefly reviewed in App.~\ref{app_TM}), which allows one to extract the kink density $n_\beta$ from Eq.~\eqref{eq_twopt_thcor}\cite{PhysRevB.11.3535,PhysRevB.22.477,PhysRevE.48.4284} and thus confirm the accuracy of Eq.~\eqref{eq_n_beta}.

Note that Eq.~\eqref{eq_twopt_thcor} enjoys a certain degree of universality. In fact, independently of the actual details of the model encoded in the specific form of the double-well potential $V(\phi)$, the two-point correlation function of the field $\phi$ at two points $x$ and $y$ assumes the simple form of an exponential, provided that $|x-y|\gg m^{-1}$, where the associated correlation length $\xi = (2 n_\beta)^{-1}$ is uniquely determined by the density of kinks $n_\beta$.
In view of this emerging degree of universality at thermal equilibrium, in Sec.~\ref{sec_OUT} we investigate the natural question whether the presence of topological excitations induce some sort of universal behavior also out of equilibrium.

\subsection{A glimpse into the quantum world}
\label{sec_quantum_effects}

In this short section, we briefly consider the quantization of the Hamiltonian \eqref{eq_H_continuum} and comment on how it affects the low-temperature thermodynamics discussed above for the classical model.
According to canonical quantization, the classical fields $\phi(x)$ and $\Pi(x)$ are promoted to operators 
$\hat{\phi}(x)$ and $\hat{\Pi}(x)$, respectively, which satisfy the commutation relation $[\hat{\phi}(x),\hat{\Pi}(y)]=i\delta(x-y)$.
As we did in the classical case, we can start by considering a single static kink centered at the origin: 
its quantization can be achieved by adding quantum fluctuations $\hat{\chi}(x)$ around the classical solution $\phi^K(x)$, i.e., by considering\cite{kinkbook} $\hat{\phi}(x)=\phi^K(x)+\hat{\chi}(x)$.
This is essentially the same decomposition as that used in~Eq.~\eqref{rad_spl} to describe radiative modes,  with the important difference that now the radiation field $\hat{\chi}$ is a quantum operator satisfying $[\hat{\chi}(x),\hat{\Pi}(y)]=i\delta(x-y)$.
The ground-state quantum fluctuations of the radiation renormalize\cite{kinkbook}  the kink mass to a value $M_R$ compared to its classical value $M$ and, also in this case, a Lorentz boost can set the kink into motion with velocity $v$ and associated energy $E=\gamma(v) M_R$.
As in Sec.~\ref{sec_thermo}, radiative modes can still be understood as fluctuations around the minima $\pm\bar{\phi}$ of the potential $V$.
Once the various renormalizations are accounted for, kinks are essentially still classical objects and therefore one can derive again Eq.~\eqref{eq_twopt_thcor} for the two-point correlator of the order parameter, under  the assumptions already discussed after Eq.~\eqref{eq_twopt_thcor}.
In equilibrium at low temperatures, the classical radiative correlator in
Eq.~\eqref{eq_chi_corr} is modified because now $\hat{\chi}$ is a bosonic quantum field obeying Bose-Einstein statistics. Indicating by $:\cdots:$ the normal ordering, one finds:
\be\label{eq_chi_corr_quantum}
\langle\,:\!\hat{\chi}(x)\hat{\chi}(y)\!:\, \rangle\simeq \int_{-\infty}^{+\infty}\!\! \frac{\dd k}{2\pi}\frac{1}{\epsilon(k)} \frac{e^{ik(x-y)}}{e^{\beta\epsilon(k)}-1},
\ee
where the energy is terms of the momentum $k$ is still given by $\epsilon(k)=\sqrt{k^2+m_R^2}$, as in the classical case, but the classical mass $m=V''(\pm\bar{\phi})$ of the fluctuation gets renormalized\cite{kinkbook} to the value $m_R$, similarly to what happens to the kink mass $M \to M_R$. 
The correlation length $\xi$ which controls the exponential decay of $\langle\,:\!\hat{\chi}(x)\hat{\chi}(y)\!:\, \rangle$ upon increasing $|x-y|$ can be extracted from the integral above and it is still given by the mass scale $\sim m_R^{-1}$. However, an important difference emerges between the classical and the quantum case:
While in the classical case Eq.~\eqref{eq_chi_corr} predicts that the amount of radiation quantified by $\langle\,:\!\hat{\chi}^2\!:\, \rangle$ is algebraically suppressed as $\sim \beta^{-1}$ at low temperatures $\beta\to\infty$, in the quantum case Eq.~\eqref{eq_chi_corr_quantum} implies an exponential suppression 
$\sim e^{-\beta m_R}$ in  the same limit.
Accordingly, while in the classical case the radiation always 
dominates over the kink density at sufficiently low temperature, in the quantum one this depends on the ratio between the two mass scales $M_R$ and $m_R$, which are determined by the details of the potentials $V(\phi)$.
As we will see in Sec.~\ref{sec_phen}, having more radiation than kinks (in the sense specified therein) has important consequences on the non-equilibrium dynamics of the system
: in practice, this requires $M_R>m_R$. Particularly interesting is the limit $M_R\gg m_R$ since, depending on the temperature, the quantum system may be well-described by the classical model \ocite{RAJARAMAN1975227,MUSSARDO2007101,doi:10.1080/09500340008232189,doi:10.1080/00018730802564254,PhysRevLett.122.120401}. Out of equilibrium, the range of validity of this semiclassical approximation is set by the energy scale, or, equivalently, the temperature attained by the system in the long-time limit after thermalization takes place.
At relatively high temperatures $\beta m_R\ll 1$ (but still $\beta M_R\gg 1$ in order to stay within the regime of low density of kinks) the semiclassical approximation holds: indeed, in this case Eq. \eqref{eq_chi_corr_quantum} can be approximated with the classical expression in Eq. \eqref{eq_chi_corr}. On the other hand, if $\beta m_R\gtrsim 1$ the system is far from being classical and quantum effects become important: however, the general mechanism leading to a time-scale separation between the kinks and the radiation dynamics is expected to be still effective, as we extensively comment.

\subsection{Effects of the lattice}
\label{sec_lattice_effects}

In general, the introduction of a finite lattice spacing $a>0$ in Eq.~\eqref{eq_H} complicates the analytical treatment compared to the continuum limit in Eq.~\eqref{eq_H_continuum}.  
On the one hand, the quantization of the classical lattice model can be done as explained in 
Sec.~\ref{sec_quantum_effects} above, namely by quantizing the radiative modes, with no additional difficulties. 
On the other hand, the breaking of translational invariance due to the underlying lattice makes the dynamics of the topological excitations extremely complicate and analytically untractable.
However, a detailed analysis can be done in the limit of small (but finite) lattice spacing by using the continuum model in Eq.~\eqref{eq_H_continuum} as the zeroth-order approximation. 
This approach is justified, however, only if the kink width $w$ is large compared to the lattice spacing $a$: fulfilling this condition depends on both the specific form of the potential $V(\phi)$ and the kink velocity $v$, which causes a Lorentz contraction $w \to w/\gamma$ of the width $w$ of a moving kink.
However, within the range of  low temperatures which we are eventually interested in, kinks are slowly moving and therefore we do not expect the latter effect to be relevant.

The presence of a lattice breaks translational invariance. As a result, a kink does no longer move at constant velocity, but it feels the effect of the so-called Peierls-Nabarro potential\cite{Nunes1940,nabarro1967theory} which has the same periodicity as the underlying lattice.
Due to this potential and the corresponding force, the kink couples with the radiative modes\cite{kinkbook} and  it emits radiation while moving, progressively losing energy and experiencing 
an effective friction\cite{PEYRARD198488,Woafo_1993,PhysRevLett.39.891,PhysRevB.33.1904}.
This potential affects significantly the behavior of a single kink compared to the continuum limit, but the emerging differences with the latter decrease at finite temperature
temperature\cite{TRULLINGER1987181,DIKANDE1994283,PhysRevB.41.4570}.
As discussed in Sec.~\ref{sec_thermo}, the thermodynamic behavior of the system on the lattice can still be understood in terms of a dilute gas of kinks moving in a background radiation but lattice corrections have to be included.  
Similarly, apart from quantitative lattice corrections, the qualitative behavior at low temperature of both the radiation and the kink density as a function of temperature are not affected compared to the model on the continuum.

\section{Out of equilibrium: emergence of universal late-time dynamics}
\label{sec_OUT}

After having described the equilibrium properties of the field theory in Eq.~\eqref{eq_H_continuum} 
induced by the topological excitations at low temperatures, we now consider the non-equilibrium behavior of the same theory, looking for a possible emergent universal behavior due the existence of kinks.
We start with a detailed description of the class of protocols used to drive the system out of equilibrium and subsequently we build upon our understanding of the thermodynamics in order to describe the long-time dynamics, which turns out to be accurately described by a simple model focussed on kinks.
The physical arguments supporting our results are actually valid for both the classical and the quantum version of the system and we conveniently benchmark them in the former via numerical simulations.

\subsection{Quench protocol}
\label{subsec_quench}
The non-equilibrium dynamics of the system is realized by means of the following quench protocol: for times $t<0$ the evolution is governed by the Hamiltonian
\be\label{eq_pre_q_H}
H_0 = a\sum_x \left[ \frac{\Pi^2(x)}{2}+\frac{(\phi(x+a)-\phi(x))^2}{2 a^2} +V_0(\phi(x))\right],
\ee
where $V_0(\phi)$ is a potential with a single minimum, assumed to be located at $\phi=0$.
We initialize the system in a steady state of the Hamiltonian $H_0$ with $\mathbb{Z}_2$ unbroken symmetry: for example, we consider an initial thermal ensemble with inverse temperature $\beta_0$. Within the quantum case, by letting $\beta_0\to\infty$, one can also select the ground state of $H_0$  as the initial pure state.
At time $t=0$ the potential is suddenly switched from $V_0(\phi)$ to $V(\phi)$ and the system is let evolve with the Hamiltonian~\eqref{eq_H} where, as in the previous Sections, the potential $V$ is invariant under  
$\mathbb{Z}_2$ symmetry $V(\phi)=V(-\phi)$ and has two degenerate minima $\pm\bar\phi$.
In the classical case, initial field configurations are randomly sampled from the thermal ensemble of $H_0$ \eqref{eq_pre_q_H} and then they are independently evolved via the dynamics generated by the deterministic Hamiltonian~\eqref{eq_H}. Observables are then computed along the time evolution and averaged over the initial conditions. 
This averaging is useful in order to reduce the impact of fine-tuned initial field configurations which might leat to  a non-generic behavior. Note that the same preparation of initial conditions was made in 
Refs.~\ocite{Nessi1,Nessi2,Nessi3} and that this protocol corresponds to a quantum quench\cite{Calabrese_2007} when referred to a quantum system.

In the absence of conservation laws beyond that of the energy,  the system is expected to eventually thermalize after the quench. 
In this respect, it is important to consider systems with a finite lattice spacing $a>0$, assumed to be anyhow smaller than the kink width $w$, see Sec.~\ref{sec_lattice_effects}. 
In fact, in the continuum limit $a\to 0$, most of the energy of the system is stored as kinetic energy of the modes with large momenta, for which interactions are largely irrelevant: accordingly,  the redistribution of energy among the various modes, driven by the interactions and necessary for thermalization, is severely hindered\cite{BoDeVe04}. 
For this reason, we focus below on lattices with finite lattice spacing, for which the general picture discussed in Sec.~\ref{sec_themodel} is valid: the kinks determine the large-distance behavior of the two-point correlator in 
Eq.~\eqref{eq_twopt_thcor} and their density decreases exponentially fast upon decreasing the temperature, while the amount of radiation does so only algebraically (in the classical case).

As discussed in Sec.~\ref{sec_thermo}, universal features emerge in equilibrium at low temperatures: in order to investigate their consequences out of equilibrium, the quench protocol has to be chosen such that the eventual stationary state of the system is characterized by a sufficiently low temperature. 
This can be done by properly selecting the ensemble of initial conditions and the parameters of the quench, i.e., those of $V_0$ and $V$.
Indeed, we emphasize that the initial and final inverse temperatures of the system, i.e., $\beta_0$ and $\beta$, respectively,  are generically different: the final temperature $\beta$ at which the system eventually thermalizes is determined by the total energy injected in the system by the quench, which depends, inter alia, on $\beta_0$ and which is conserved by the dynamics separately for each specified initial configuration of the field.
In practice, the final inverse temperature $\beta$ is implicitly determined by the equality
$\langle H \rangle_\beta = \langle H \rangle_{\beta_0}$ where $\langle \cdots \rangle_{\beta,\beta_0}$ are the final and initial thermal ensembles, respectively, i.e., 
\be\label{eq_energy_con}
\frac{\int [\dd \phi] H[\phi] e^{-\beta H[\phi]}}{\int [\dd \phi] e^{-\beta H[\phi]}}=\frac{\int [\dd \phi] H[\phi]e^{-\beta_0 H_0[\phi]}}{\int [\dd \phi] e^{-\beta_0 H_0[\phi]}}\, .
\ee
This equation can be solved by using the fact that $H_0$ on the r.h.s.~is Gaussian and by evaluating efficiently the l.h.s.~with the  transfer-matrix algorithm discussed in Appendix \ref{app_TM}.
In practice, we consider the prequench Hamiltonian in Eq.~\eqref{eq_pre_q_H} for a finite system of length $L$,  lattice spacing $a=1$, and with the harmonic potential $V_0(\phi)=m_0^2 \phi^2/2$. 
The length $L$ is chosen to be sufficiently large for the system to approximate its thermodynamic limit, i.e., $L$ has to be larger than any other macroscopic and mesoscopic scale in the system, e.g., one should require $L n_\beta\gg 1$.
The values of the initial mass $m_0$ and of the inverse temperature $\beta_0$ are the free parameters which we use to change the initial conditions, being always careful to remain in the low-temperature regime for final thermal states.
In view of the emerging universal behavior discussed in the Sections which follow,  we do not expect that different choices of the parameters of the initial conditions or of the post quench potential will affect our conclusions as long as the requirements mentioned above are fulfilled.  
The real-time simulations presented below are performed by using the method briefly discussed in Appendix \ref{app_Met}, while thermal expectation values are computed with the transfer matrix approach described in Appendix \ref{app_TM}.

\subsection{Phenomenological description}
\label{sec_phen}

Since the initial states of the evolution of the system are described by a $\mathbb{Z}_2$-symmetric statistical ensemble at low temperature, the corresponding field configurations consist of small fluctuations around the configuration with $\phi=0$.
When the initial quadratic potential $V_0(\phi)$ is changed at time $t=0$ into $V(\phi)$, the configuration $\phi=0$ corresponds to a local maximum of $V(\phi)$ which it is no longer stable under the temporal evolution: accordingly, the field locally tends to approach the values  $\pm\bar\phi$  corresponding to the two possible degenerate vacua of $V(\phi)$.

The initial fluctuations will determine towards which of these two values the field $\phi$ locally evolves, generically resulting into configurations in which, for a fixed time $t$, $\phi(t,x)$ alternates between 
$\pm\bar\phi$ as a function of $x$: a kink is present at the position at which the field switches sign.

\begin{figure}[t!!]
\includegraphics[width=0.95\columnwidth]{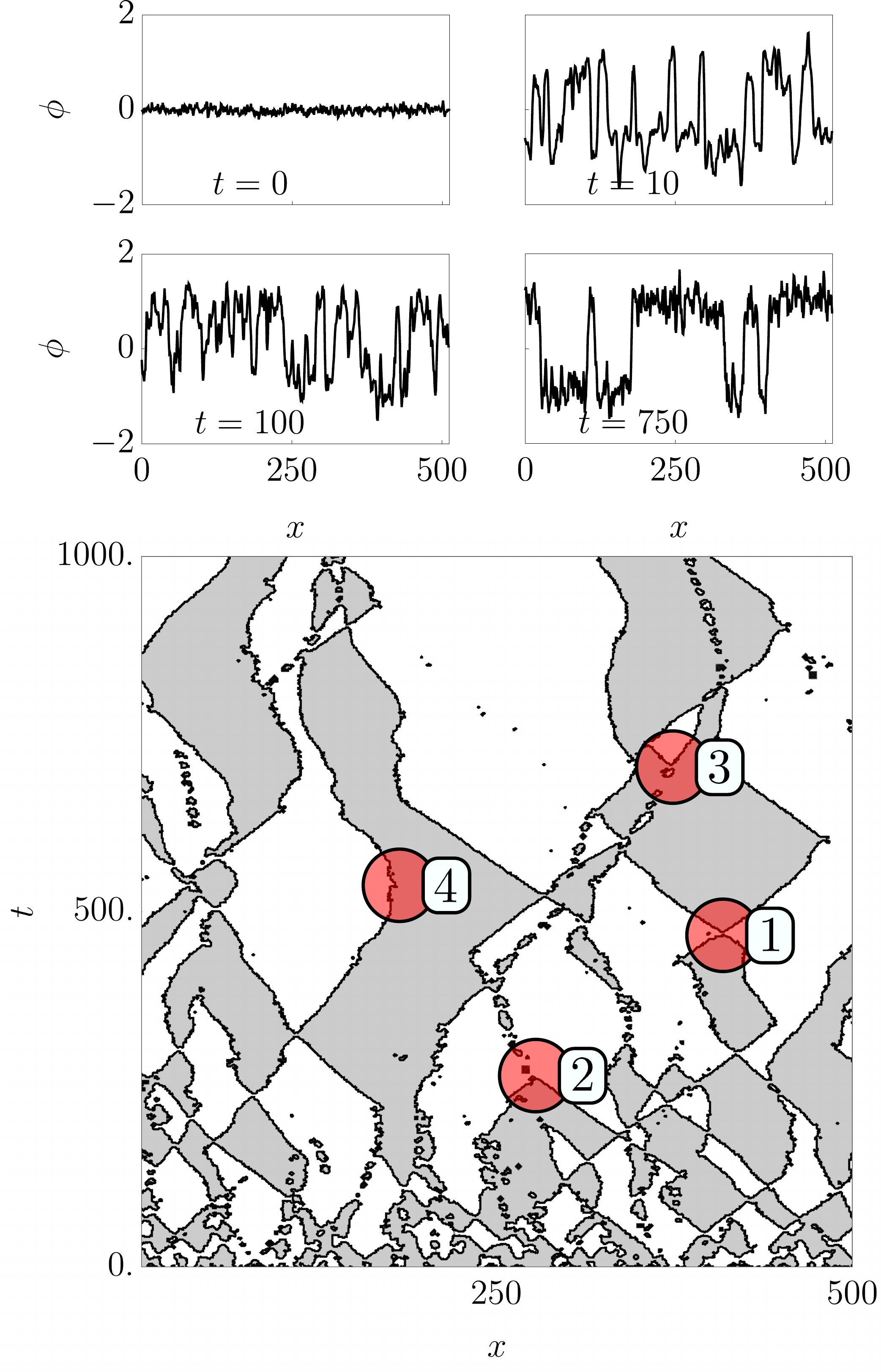}
\caption{\label{fig_4} 
(Color online) Upper panel: $\phi(t,x)$  as a function of $x$ at various times $t$, evolving from the initial field configuration represented on the top-left plot. The initial small fluctuations grow in amplitude, the field $\phi(t,x)$ locally approaches $\pm\bar\phi$ and a sequence of alternating kink and antikinks --- located where $\phi(t,x)$ vanishes --- emerges at a coarse-grained scale, i.e., neglecting the fluctuations due to radiation. 
Lower panel: trajectories of the kinks in the space-time $(x,t)$-diagram, starting from the same initial condition as above. Regions with $\phi(x)<0$ (resp. $\phi(x)>0$) are colored in gray (resp. white) with the (anti)kinks located at the interfaces between the white and gray regions. The (approximately) ballistic nature of the motion of the (anti)kinks clearly emerge from the patterns in the plot. 
Red circles highlight the various scattering processes occurring during the evolution and described in the main text. 
The initial configuration is sampled from a thermal state of the Hamiltonian in Eq.~\eqref{eq_pre_q_H} with $V_0(\phi)=m_0^2 \phi^2/2$, $a=1$ and inverse temperature $\beta_0= 10^{2}$, while the field evolves with interaction $V(\phi)=0.1(\phi^2-1)^2$. 
The relatively small value of the chain length $L=500$ is used here only for the purpose of illustration, while the large-scale simulations discussed further below are performed on much larger systems (typically $L\simeq 2^{11}$).}
\end{figure}


Let us denote by $n(t)$ the density of kinks at a given time $t$, averaged over the initial conditions.
The value of $n(0)$ can be estimated in terms of the correlation length $\xi_0$ of the initial state: in a first approximation, portions of the system with a size smaller than $\xi_0$ behave as an unique entity and choose one of the two vacua $\pm\bar\phi$, while points further away are independent and therefore free to choose different values, creating a kink in the middle.
As discussed in Sec.~\ref{sec_thermo} (with the replacement $m\to m_0$ and $\beta\to\beta_0$ in Eq. \eqref{eq_chi_corr}), 
$\xi_0$ is proportional to $m_0^{-1}$,  the inverse  of the bare mass $m_0$ which characterizes the potential $V_0$. 
If $m_0^{-1}$ is larger than the typical width $w$ of the kink, we can estimate $n(0)\simeq m_0^{-1}$ and therefore the initial kink density is expected to be approximately independent of the initial average energy and of the final temperature.
Since at infinite time the system approaches a thermal state (with the final inverse temperature $\beta$ determined by the energy conservation), the kink density $n(t)$ is expected to approach its thermal value, i.e., $\lim_{t\to \infty} n(t)=n_\beta$. 
At sufficiently low temperatures, generally one has $n_\beta \ll n(0)$ and therefore $n(t)$ must diminish along the time evolution.

In Fig.~\ref{fig_4} (upper panel) we show different snapshots of the field profile $\phi(t,x)$ as a function of $x$ at various times $t$, for a single random initial condition $\phi(t=0,x)$. 
We can observe the kinks clearly emerging from the radiative background at $t=0$ and their spatial density progressively reducing, as the time evolution proceeds further.
After their creation, kinks start traveling across the system, interacting with each other and with the background radiation. In Fig.~\ref{fig_4} (lower panel) we highlight the trajectories of the (anti)kinks associated with the field configuration presented also in the upper panel of the same figure and corresponding to the boundaries of the grey  regions in the $(x,t)$-diagram.  Most of the time, the (anti)kinks behave as free particles with a well-defined velocity, traveling along straight lines in that diagram. However, in addition to this free motion, various scattering processes take place, as highlighted with the numbered red circles in Fig.~\ref{fig_4} (lower panel):
\vspace{-0.25cm}
\begin{itemize}
\item[(1)] A kink and an antikink  scatter elastically;
\vspace{-0.25cm}
\item[(2)] A kink annihilate with an antikink;
\vspace{-0.25cm}
\item[(3)] A kink-antikink pair is created from the fluctuations background radiation;
\vspace{-0.25cm}
\item[(4)] A single kink scatters against the background radiation and suddenly changes the direction of its motion. 
\end{itemize}
\vspace{-0.25cm}
Universal behavior is generically not expected to emerge at short times because the dynamics of the system is still significantly affected by the disordered initial condition. In contrast, a universal approach to the thermal value is expected (at least at low temperatures), in view of the arguments presented further below. 

\subsubsection{Near the thermal state}
In order to develop our intuition, we start by focussing on the features of the eventual equilibrium thermal state of the evolution. As shown by Eq.~\eqref{eq_chi_corr}
in the classical case, the amount of radiation $\langle \chi^2\rangle$ vanishes $\propto \beta^{-1}$ as a function of the temperature $\beta^{-1}$ upon decreasing it,  while the density $n_\beta$ of kinks vanishes exponentially fast in the same limit, as in Eq.~\eqref{eq_n_beta}.
As we discussed in Sec.~\ref{sec_quantum_effects}, in the quantum case both the amount of radiation and the density of kinks vanish exponentially fast as functions of $\beta^{-1}$ at small temperature: however, if the mass $M$ of the kink is larger than that of the radiation, at low temperatures we will eventually be in a situation
in which there is more radiation than kinks.
Accordingly, as a rule of thumb, the process which occurs more frequently is the self-interaction of the radiation.

Now, let us slightly perturb the system, driving it out of equilibrium in such a way that it is still close to the thermal ensemble. 
The actual details of this perturbation do not really matter because, as long as the system remains in the low-temperature regime, 
we argue that the time-scale separation between the (slow) dynamics of the kinks and the (fast) one of the radiation persists and it is responsible for the emergence of a largely universal behavior.

The radiation is expected to  thermalize due to the anharmonic corrections in Eq.~\eqref{eq_rad}. In the classical case, the timescale $\tau^\text{rad}_\text{th}$ for this thermalization has been found to depend on the initial energy density $\epsilon$, assumed to be small, as $\tau^\text{rad}_\text{th}\propto \epsilon^{-\gamma}$ with $\gamma>0$, in the case of a single-well potential\ocite{BoDeVe04}. In the presence of a potential with a double well, for most of the time the radiation fluctuates around one of the two vacua, as it does in the presence of a single well. Accordingly, we can estimate the relaxation time scale of the radiation in terms of its typical energy scale which, in view of the quadratic approximation in Eq. \eqref{eq_chi_corr} for which equipartition holds, is expected to be proportional to the temperature $\propto \beta^{-1}$. 
Now, let us consider the kink annihilation processes. Since a single kink in the radiative background is stable, 
annihilation can occur only when two kinks meet at the same point. The probability for a single kink to be annihilated  is thus proportional to the number of kinks it meets along its trajectory, i.e., $\propto n_\beta$. Accordingly, we get a rough estimate of the time scale $\tau_\text{th}^\text{k}$ associated with the relaxation of kinks, i.e., $\tau_\text{th}^\text{k}\sim n_\beta^{-1}$. 
Note that, while $\tau^\text{rad}_\text{th}$ algebraically grows for large $\beta$ upon increasing it, the timescale $\tau_\text{th}^\text{k}$ correspondingly grows exponentially: for sufficiently low temperature temperatures this imply a clear separation between these two time scales.

The qualitative picture of the dynamics emerging from the facts mentioned above is that of a collection of kinks moving (almost ballistically) in a radiative background, which thermalizes on a time scale much shorter than that of the kinks. 
In addition, we argue that there is a further time-scale separation between the total density $n(t)$ of topological excitations and the other degrees of freedom related to kinks, such as the kink velocities. 
This hypothesis is based on two observations:
First, above we estimated $\tau_\text{th}^\text{k}$ considering scattering events between kinks which result  in  a variation of their number but, for most of the time,  kinks travel across the radiative background as isolated particles. In this case, the radiation cannot destroy a single kink, but it affects its velocity which is then expected to relax faster.
Second, scattering among kinks hardly results into annihilation: in fact, this would require converting the large amount of energy stored in the kinks (at least twice the kink mass $M$) into radiation, which typically has an energy $\beta^{-1}$. 
Looking at the bottom panel of Fig.~\ref{fig_4} we see that, indeed, kinks survive to most of the scattering they undergo along their trajectories. 
However, while kink annihilation rarely occurs, inelastic collisions in which the momenta of the incoming kinks are slightly changed after the scattering are much more frequent: these processes require only a small energy exchange with the background radiation and contribute to the scrambling of the kink momenta.
Accordingly, we expect the velocity of a kink to thermalize much faster than the timescale on which the total density is affected.

\subsubsection{%
Dynamics far from equilibrium: long-time behavior
}

We now provide numerical support the hypothesis discussed in the previous Section on the emergence of the time-scale separation, working out its consequences for the dynamics of observables. 
The numerical calculations presented further below refer to the Hamiltonian in Eq.~\eqref{eq_H} with a symmetric pre-quench potential as in Eq.~\eqref{eq_pre_q_H}  and a double-well post-quench  $V(\phi)=\lambda (\phi^2-\bar{\phi}^2)^2$. 
In both the pre- and post-quench Hamiltonian, we consider the model on a lattice with $a=1$.
The values of $\lambda$ and $\bar\phi$ used in the numerical analysis are chosen considering the following competing trends. First note that, for a fixed $\bar{\phi}$, $\lambda$ controls the height of the barrier separating the two vacua and therefore the mass of the kink $M$. Using expressions valid on the continuum as estimates of the corresponding ones on the lattice, Eq.~\eqref{eq_rest_mass} implies that a small value of $\lambda$ results into a small kink mass $M$.
In turn, according to Eq.~\eqref{eq_n_beta}, a small kink mass $M$ requires larger temperatures in order to achieve the limit of small kink density, at which a universal non-equilibrium behavior is expected to emerge.
In the opposite case of large $\lambda$, the kink width $w$ obtained from Eq.~\eqref{eq_kink_phi4} decreases and lattice effects become predominant, as discussed in Sec.~\ref{sec_lattice_effects}). 
Based on our numerical experience, the choice $\bar{\phi}=1$ and $\lambda=0.1$ allows us to study the phenomena under investigation within space-time scales which are numerical accessible to our calculations. 
Accordingly, the numerical data discussed further below are provided for this specific choice of couplings and for various initial conditions.

Let us start by considering the evolution of the kink density $n(t)$. 
At short times, $n(t)$ is large and the system is far from the equilibrium state. 
However, $n(t)$ decreases upon increasing $t$: the conditions under which Eq.~\eqref{eq_corr_ord} was  derived are eventually satisfied and therefore we expect the equal-time correlator of the order parameter to obey an analogous expression, i.e.,
\be\label{eq_correlation_time}
\langle \phi(t,x)\phi(t,y)\rangle= A(t) \, e^{-2n(t)|x-y|},
\ee
for $|x-y|$ much larger than the typical correlation length of the radiation. 
The prefactor $A(t)$ in Eq.~\eqref{eq_correlation_time} is not universal and it is due to the renormalization induced by the evolving background radiation. 

\begin{figure}[t!]
\includegraphics[width=1\columnwidth]{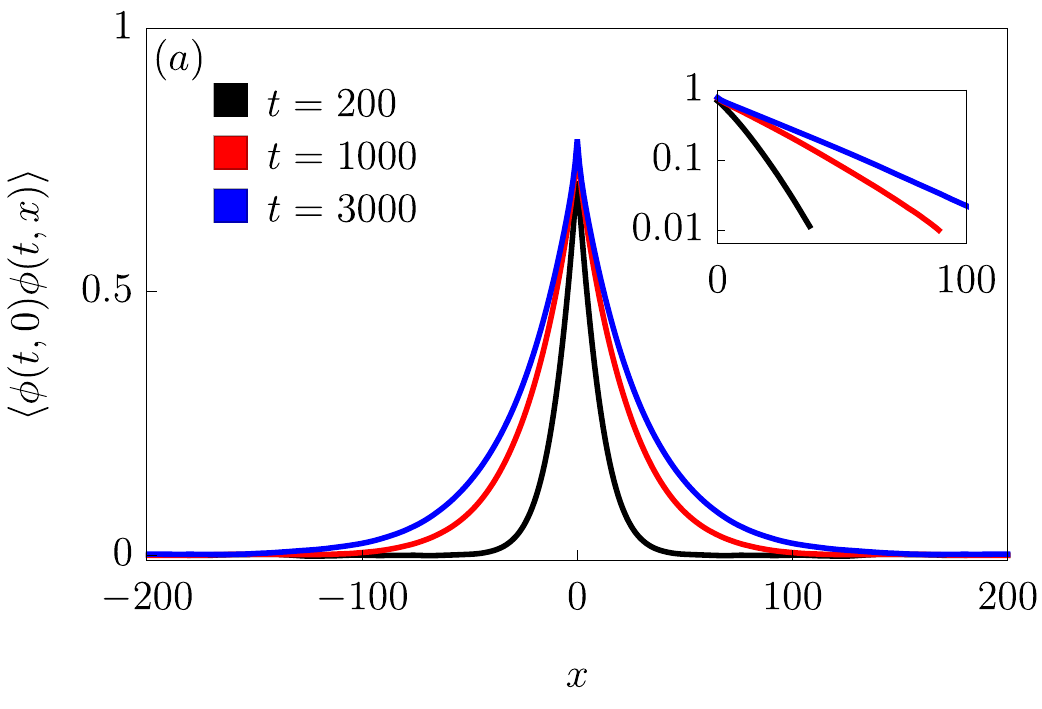}\\ 
\includegraphics[width=1\columnwidth]{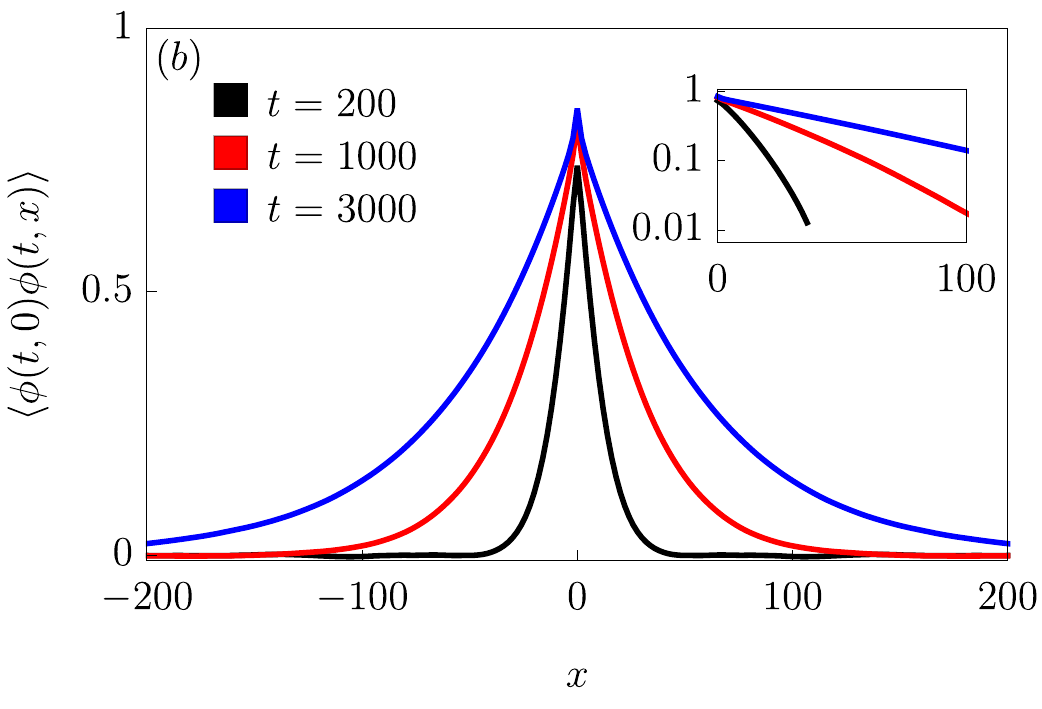}
\caption{\label{fig_5} (Color online) Order parameter correlation function $\langle \phi(t,0)\phi(t,x)\rangle$ after a quench from the symmetric to the symmetry-broken potential as a function of distance $x$, 
for various times $t$. Panel (a) corresponds to an initial temperature $\beta_0=18.2$ and a final one $\beta=8$, while panel (b) to an initial temperature $\beta_0=107.2$ and a final one $\beta=10$. 
Insets: the same correlation functions are shown on a linear-log scale emphasizing the exponential decay which extends to larger distances as time increases, i.e., as $n(t)$ further diminishes.
In both panels $a=1$, $V(\phi)=0.1(\phi^2-1)^2$, $m_0=1$, while the actual value of $L$ was chosen depending on each final temperature as the most convenient one in order to satisfy the condition discussed at the end of Sec.~\ref{subsec_quench}, see also Appendix~\ref{app_Met}.
}
\end{figure}

Equation~\eqref{eq_correlation_time} is expected to apply as soon as the system enters the regime of low-densities regime, still far from equilibrium because $n(t)\gg n_\beta$. 
Accordingly, Eq.~\eqref{eq_correlation_time} is expected to apply 
at times smaller than time scale over which global thermalization occurs. 
In Fig.~\ref{fig_5} we show the large-distance behavior of the correlation function $\langle \phi(t,0)\phi(t,x)\rangle$ at various times $t$ and for different initial conditions, which result in different final temperatures of the eventual thermal state.  The exponential decay predicted by Eq.~\eqref{eq_correlation_time} is reproduced with good agreement, as highlighted by the insets and the agreement increases as time increases, since $n(t)$ decreases.
As we have already anticipated in Sec. \ref{sec_thermo}, a direct counting of the kinks is difficult, especially at relatively high temperatures, since large radiative fluctuations can be wrongly identified with an overlapping kink-antikink pair. Accordingly, we use the slope of the exponential decay in Eq.\eqref{eq_correlation_time} in order to determine numerically the value of $n(t)$.
 
At later times, as the kink density $n(t)$ decreases, the system eventually enters a regime within which the kink annihilation processes take place on a much longer timescale compared to that controlling the relaxation of the radiation. Accordingly, the radiation can be thought of as to be instantaneously in a thermal ensemble with an effective temperature which depends on time. 
Indeed, as kinks annihilate, their energy is converted into radiative modes, the temperature of which slightly changes as a consequence of the extra energy that they receive.  
Note, however, that in the limit of low kink density the energy which is transferred to radiative modes is also negligible and therefore, in first approximation, one can assume the radiative modes to be effectively thermalized to the final temperature that the system will attain after the occurrence of the eventual relaxation.
%
\begin{figure}[t!]
\includegraphics[width=0.7\columnwidth]{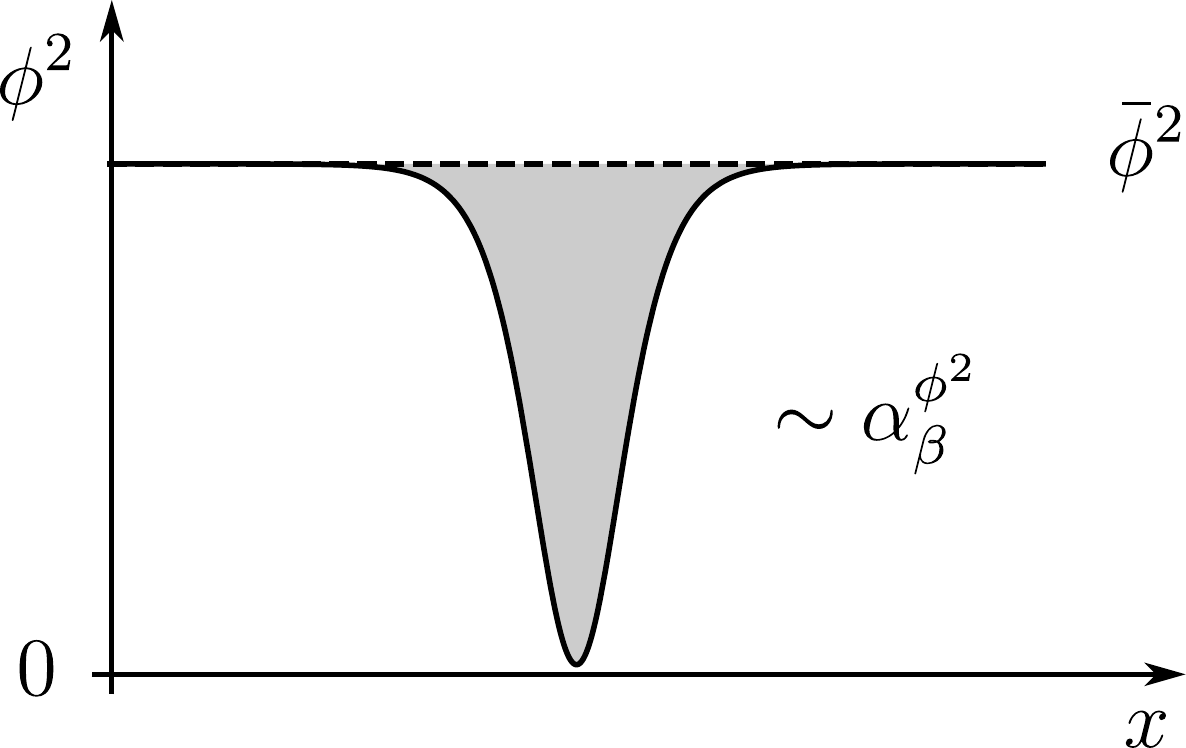}
\caption{\label{fig_6} (Color online) Cartoon representation of Eq.~\eqref{eq_onept_nt_lin} in the specific case $O(t,x) =\phi^2(t,x) $. The space-dependence of $\phi^2(x)$ is shown when evaluated on a configuration without (dashed line) or with (solid line) a kink. The difference of the corresponding values of the integral $\int \!\dd x\,\phi^2(x)$ is represented by the shaded area, which quantifies $\alpha_\beta^{\phi^2}$ for a single kink.  When kinks are sufficiently diluted, each of them independently contributes to this difference, resulting into Eq.~\eqref{eq_onept_nt_lin}.}
\end{figure}
%
We now demonstrate that $\mathbb{Z}_2$-even observables are also sensitive to topological excitations and therefore they provide another useful tool to investigate their dynamics. 
Let us consider a $\mathbb{Z}_2$-even local observable $O(t,x)$, for example the even powers of the field $\phi^{2n}(t,x)$.
Under the hypothesis that all degrees of freedom relax quickly after the quench with the sole exception of the kink density $n(t)$, the expectation value of $O$ must be a function of the latter, i.e.,
\be\label{eq_onept_nt}
\langle O(t,x)\rangle= \overline{O}[n(t)].
\ee
We know that at infinite time $\langle O(t,x)\rangle$ and the kink density $n(t)$ must reach their thermal values, i.e., $\overline{O}_\beta$ and $n_\beta$, respectively.
Accordingly, at sufficiently low kink densities $n(t)$ one can expand Eq.~\eqref{eq_onept_nt} in powers of $n(t)$ as
\be\label{eq_onept_nt_lin}
\langle O(t,x)\rangle=\overline{O}_\beta+\alpha^O_\beta (n(t)-n_\beta)+\mathcal{O}(n^2(t)), 
\ee
where $\alpha^O_\beta$ is an observable-dependent non-universal constant determined by the equilibrium properties at the final temperature. The term of the expansion of the zeroth order in $n(t)$ is fixed by the requirement that in the long-time limit (where $n(t)\to n_\beta$) the observable attains its thermal value.
Note that, in order for Eq.~\eqref{eq_onept_nt_lin} to be valid, two independent conditions have to be satisfied.
First, Eq.~\eqref{eq_onept_nt} is assumed to be valid, i.e., all the degrees of freedom have relaxed except for the average kink density.
Second, the kink density is assumed to be sufficiently small to allow a linearization of Eq.~\eqref{eq_onept_nt}: this requires $n(t)$ to be small, but does not necessarily imply that $n(t)$ must be of the same order as $n_\beta$.
This condition is the same under which Eq.~\eqref{eq_correlation_time} is valid, but 
Eq.~\eqref{eq_onept_nt_lin} relies on the stronger assumption that any degree of freedom except the total density of kinks has relaxed.
%
\begin{figure}[b!]
\includegraphics[width=0.99\columnwidth]{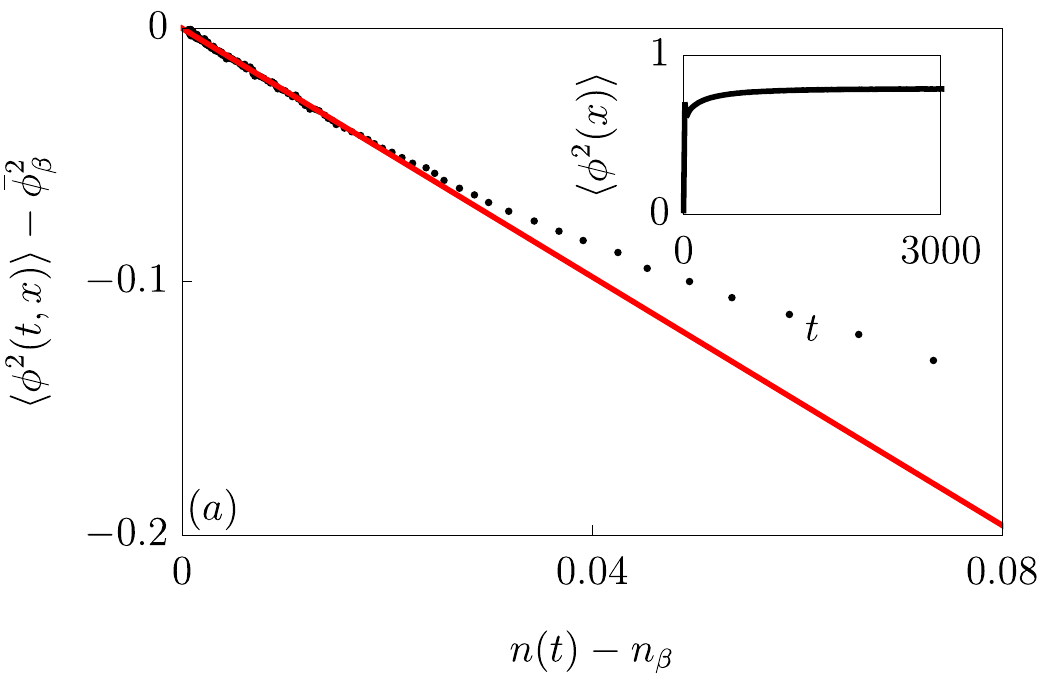}\\
\includegraphics[width=0.99\columnwidth]{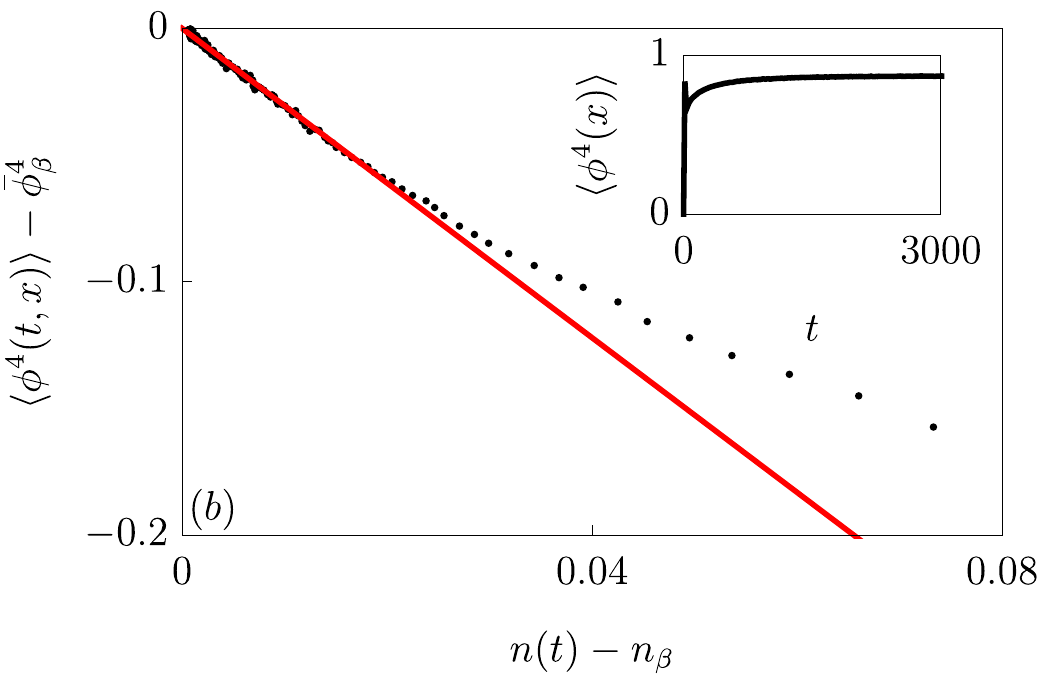}
\caption{\label{fig_7} (Color online) Parametric plot of the time evolution of $\langle\phi^2\rangle$ (upper panel) and $\langle \phi^4\rangle$ (lower panel) versus the instantaneous kink density $n(t)$, as obtained from numerical calculations (data points); the red line is a guide to the eye and indicated a linear dependence as the one predicted by Eq.~\eqref{eq_onept_nt_lin}. Insets: data for the time evolution of the local observable $\langle\phi^2\rangle$  and $\langle \phi^4\rangle$, used in the main plots.
The numerical calculations have been done for a chain with lattice spacing $a=1$, post-quench potential $V(\phi)=0.1(\phi^2-1)^2$, pre-quench thermal ensemble with potential $V_0(\phi) = m_0^2\phi^2/2$, mass $m_0=1$, and initial temperature $\beta_0=18.2$, leading to a final temperature $\beta=8$ and kink's density $n_\beta=1.6\times 10^{-2}$.
The chain length $L$ is chosen here as indicated in Fig. \ref{fig_5}. 
}
\end{figure}

The physical origin of $\alpha_\beta^O$ can be understood as follows. 
For concreteness, we focus on the simplest case $O(t,x) = \phi^2(t,x)$, but the line of argument readily extends to generic $\mathbb{Z}_2$-even local observables $O$. Consider a field configuration with a single kink and plot the profile of the observable $O$, as done in Fig.~\ref{fig_6}. This profile will mostly fluctuate around $\bar{\phi}^2$ except in the spatial region close to the location of the kink, where the value of $\phi^2(t,x)$ is depleted. Thus the spatial average of $\phi^2(t,x)$ is given approximately by $\bar{\phi}^2$ minus the contribution of the kink, which vanishes if the kink is removed. 
This phenomenological argument focuses on a single kink: considering now the infinite system with a certain density of kinks, the same argument can be independently applied to each kink on the line. Hence, the difference between the average of $O$ and its thermal value is just proportional to the difference between the instantaneous kink density and the thermal one, which is essentially the statement of Eq.~\eqref{eq_onept_nt_lin}. 
In this simple argument, we neglected the effect of the radiation, but converting a kink into radiative modes affects the temperature of the latter which then changes the renormalization of the plateau: the argument can be suitably modified to take this fact into account.
In Fig.~\ref{fig_7}  we present numerical evidence for the validity of  Eq.~\eqref{eq_onept_nt_lin} by plotting the l.h.s.~of that equation for the specific cases $O(t,x) = \phi^2(t,x), \phi^4(t,x)$ versus $n(t)-n_\beta$, upon varying the time $t$ elapsed from the quench. 
As shown in the insets of the plots, $\langle O(t,x)\rangle$ tends to $\overline{O}_\beta$ as $t$ increases and the corresponding data points reported in the main plots approach with great accuracy the linear dependence indicated by the red lines and predicted by Eq.~\eqref{eq_onept_nt_lin}.

\subsubsection{Kinetic equation for the kink density
}
\label{sec_kineq}

In the previous Sections we have argued and provided evidence of the fact that the total density of kinks $n(t)$ is the only degree of freedom which evolves slowly at long times. However, we have still to analyze the specific form of this evolution. 
In this Section we show that $n(t)$ obeys a kinetic equation which is consistent with the picture of the dynamics presented so far. 
Since the relaxation of $n(t)$ is expected to occur on a long time scale, it is reasonable to suppose that it can be described by a first-order differential equation 
\be
\partial_t n(t)=F(n(t)),
\label{eq:dyn_n}
\ee 
where $F$ is a function to be determined. Higher-order time derivatives of $n(t)$, being suppressed by increasing powers of the typical relaxation time scale compared to $\partial_t n(t)$, are expected to give only small corrections and therefore we neglect them.
The function $F$ must depend on the instantaneous kink density and on the final thermal state.
Within the regime of low densities, we can approximate $F(n)$ by a power series $F(n)=\sum_{j=0}^\infty c_j n^j$, 
which, in practice, can be truncated at some low order. 
The constants $c_j$ depend on the final thermal state and correspond to specific physical processes.
For example, annihilation processes in which a kink and an antikink scatter and decay into radiation, are described by a term $\propto n^2$ in the expansion of $F$, i.e., $F(n)=-\sigma_\beta n^2+\ldots$, where $\sigma_\beta>0$ is a decay rate which can be affected by the temperature of the final state.
\begin{figure}[t!]
\includegraphics[width=0.8\columnwidth]{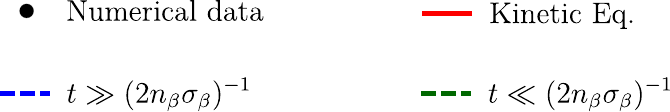}\\ 
\includegraphics[width=1\columnwidth]{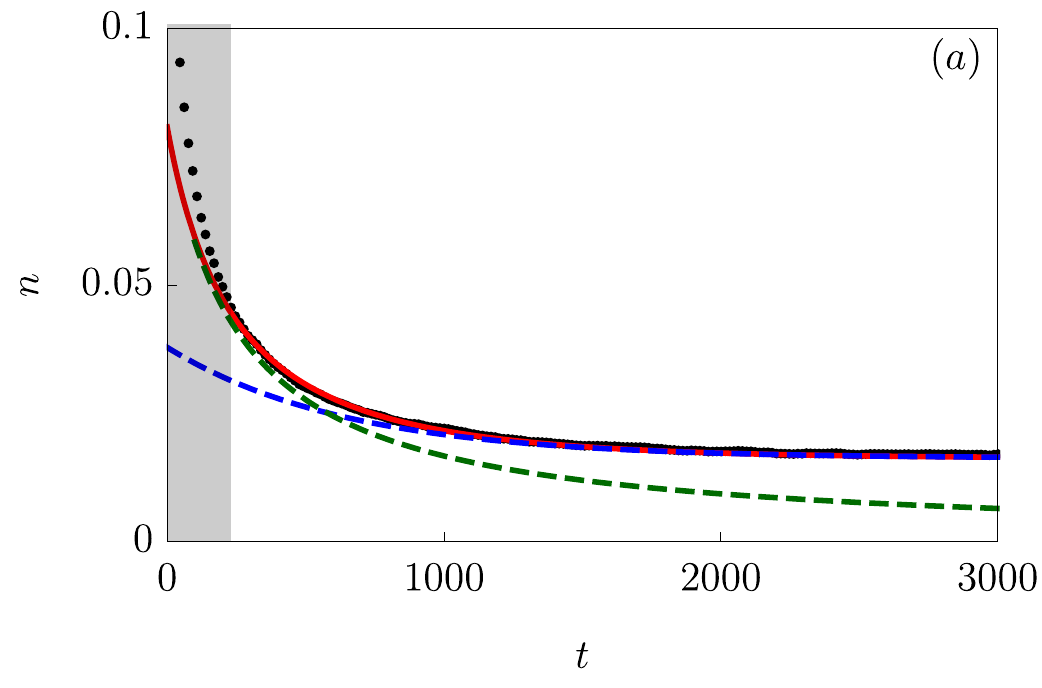}\\
\includegraphics[width=1\columnwidth]{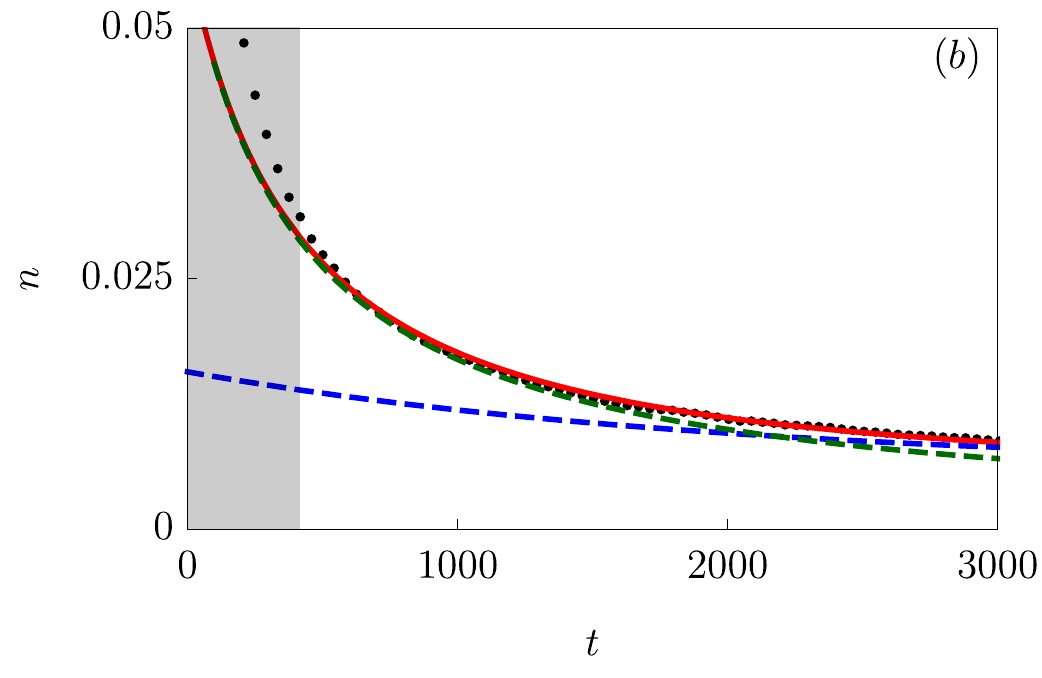}
\caption{\label{fig_8}
(Color online) Kink density $n(t)$ as a function of the time $t$ elapsed from the quench.
Data points indicate the kink density extracted from the numerical data for the two-point correlation function of the order parameter. The red solid line corresponds to the solution of the kinetic equation \eqref{eq_kin_sol} in which the (unknown) parameters $\sigma_\beta$ and $n_0$ are fixed from a best fit to the data, while $n_\beta$ is exactly computed with the transfer matrix approach.
The green dashed line indicates the short-time expansion in Eq.~\eqref{eq_power_law}, while the blue dashed line refers to the long-time behavior in Eq.~\eqref{eq_exp}. 
The kinetic equation \eqref{eq_kin_sol} is not expected to be valid at times shorter than the radiation relaxation time scale, which is indicated by the gray shaded area in the figure. This time scale can be estimated from Fig.~\ref{fig_7} as the time after which data points agree with the linear dependence indicated by the solid line. %
The numerical data reported in both figures were obtained with $a=1$, post-quench potential $V(\phi)=0.1(\phi^2-1)^2$ and an initial thermal ensemble with mass $m_0=1$ 
and inverse temperature $\beta_0$, with (a) $\beta_0=18.2$, corresponding to the final inverse temperature $\beta=8$ and (b) $\beta_0=107.2$, corresponding $\beta=10$.
The chain length $L$ is chosen here as indicated in Fig. \ref{fig_5}. 
}
\end{figure}
%
In addition to the term $\propto n^2$ identified above, one needs to include also (lower-order) terms in the expansion of $F(n)$ in order for the solution of Eq.~\eqref{eq:dyn_n} to converge 
to the expected result, i.e., $\lim_{t\to\infty}n(t)=n_\beta$.
This requires accounting for a term $\propto n^0$ which actually corresponds to the fact that, 
together with the kink annihilation processes already taken into account, 
an outburst of radiation can create a kink-antikink pair, increasing $n(t)$. 
This process depends only on the radiation and not on the actual kink density and therefore it contributes with a constant to $F$. In fact, in equilibrium, the annihilation of kinks must be balanced by these ``nucleation'' processes from radiation, such that $F(n=n_\beta)=0$.
Within the low-density regime, we can reasonably neglect terms involving powers of $n$ larger than two in the expansion of $F(n)$, but one could wonder whether a term $\propto n$ should be accounted for. 
In general, such a term is not expected: in fact, as anticipated in Sec.~\ref{sec_themodel}, 
a single (anti)kink moving into the radiation background is stable and cannot decay. Conversely, from the background radiation only kink-antikink pairs can emerge. 
Accordingly, the only way in which a $\propto n(t)$ contribution could appear is because a non-trivial correlation among kinks exists.  

Since we are neglecting these correlations in Eq.~\eqref{eq_correlation_time} we will do the same here, checking the validity of our assumptions via numerical calculations.

According to the discussion above, one can expand $F(n)$ around $n=0$ up to the second order and, by using the fact that $F(n=n_\beta)=0$, Eq.~\eqref{eq:dyn_n} can be written as
\be\label{eq_kin}
\partial_t n(t)=-\sigma_\beta[n^2(t)-n^2_\beta].
\ee
This equation can be easily solved, yielding
\be\label{eq_kin_sol}
n(t)=n_\beta +\frac{2 n_\beta}{R \,e^{2n_\beta \sigma_\beta t}-1} , 
\ee
where $R=(n_0+n_\beta)/(n_0-n_\beta)$ and $n_0$ is a parameter which depends on the initial conditions 
(see, c.f., Fig.~\ref{fig_9}(d)).
One could be tempted to evaluate Eq.~\eqref{eq_kin} at the initial time $t=0$, erroneously concluding that 
$n(0)=n_0$: however, this is not correct, because this equation is not expected to be valid at short times, but only after the radiative modes have relaxed (i.e., when Eq.~\eqref{eq_onept_nt} starts to be valid).

The solution in Eq.~\eqref{eq_kin_sol} displays two different regimes: at short times $t\ll (2n_\beta \sigma_\beta)^{-1}$ (with $t$ anyhow larger than the relaxation timescale of the radiation) it becomes
\be
\label{eq:n1_app}
n(t)\simeq n_\beta +\frac{2 n_\beta}{R \, 2n_\beta \sigma_\beta t+R-1}.
\ee
At low temperatures, $n_\beta$ is very small and therefore we typically expect $n_0\gg n_\beta$. 
Neglecting $n_\beta$, Eq.~\eqref{eq:n1_app} can be further simplified as
\be\label{eq_power_law}
n(t)\simeq \frac{n_0}{n_0\sigma_\beta t+1},
\ee
which displays an algebraic decay upon increasing time, at long times $ (n_0\sigma_\beta)^{-1}\ll t\ll (2n_\beta \sigma_\beta)^{-1}$. 
This algebraic decay causes a corresponding linear increase as a function of time of the correlation length $\propto n^{-1}(t) \propto t$ of the two-point correlation function of the order parameter in Eq.~\eqref{eq_correlation_time}. 
This growth of the correlation length can be interpreted as an instance of \emph{coarsening dynamics} \cite{Hohenberg1977,Tauber_book,Henkel_book}, as discussed further below, which is however \emph{interrupted} at longer times. 
In fact, at long times $t\gg (2n_\beta \sigma_\beta)^{-1}$,  Eq.~\eqref{eq_kin_sol} predicts an exponential relaxation of $n(t)$ to the (finite but small) thermal value $n_\beta$ upon increasing time, i.e.,
\be\label{eq_exp}
n(t)=n_\beta +2n_\beta R^{-1}\, e^{-2n_\beta \sigma_\beta t}+\mathcal{O}((e^{-2n_\beta \sigma_\beta t})^2);
\ee
correspondingly, after the linear growth discussed above, the correlation length saturates to a finite (but large) value $\propto n_\beta^{-1}$, interrupting coarsening. 

In order to test the validity of Eq~\eqref{eq_kin}, in Fig.~\ref{fig_8} we compare $n(t)$ obtained from its solution in Eq.~\eqref{eq_kin_sol} with the value obtained from the numerical analysis of the real-time dynamics of the chain. In particular, the values of the parameters $\sigma_\beta$ and $n_0$ appearing in Eq.~\eqref{eq_kin_sol} are determined from a best fit to the numerical data, while the value of $n_\beta$ is determined via the transfer-matrix approach discussed in Appendix~\ref{app_TM}. 

The agreement between the predictions of Eq.~\eqref{eq_kin} (solid red lines) and the numerical data reported in Fig.~\ref{fig_8} turns out to be excellent at long times, as expected.

%
\begin{figure}[t!]
\includegraphics[width=0.99\columnwidth]{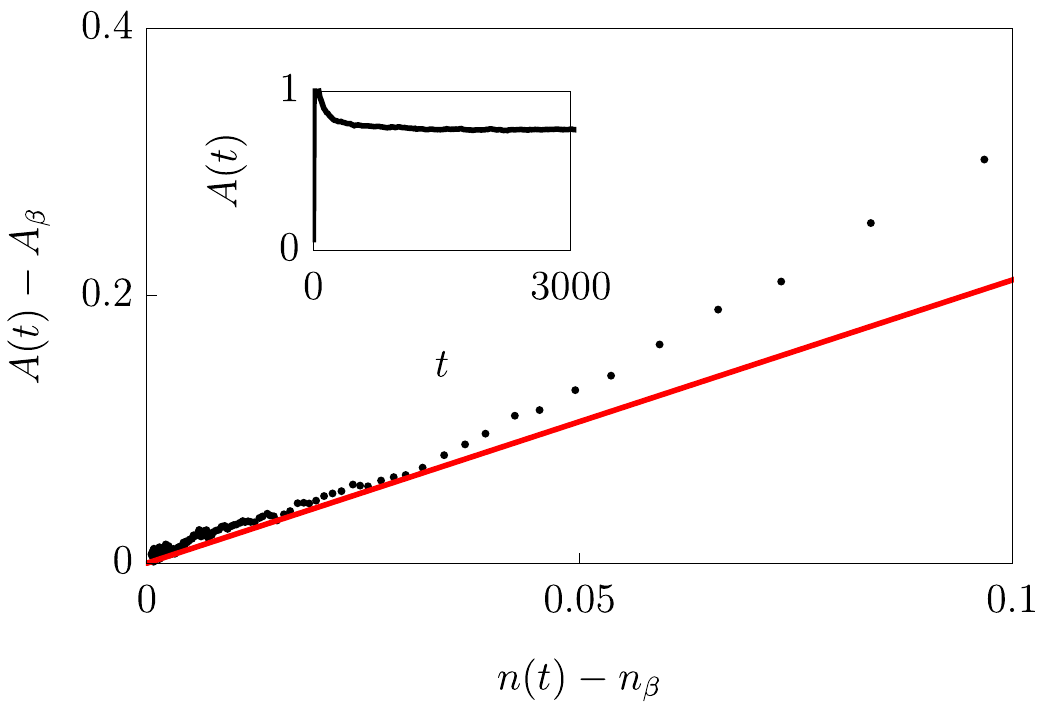}\\
\caption{\label{fig_A_tampl} (Color online) 
Parametric plot of the time evolution of $A(t)$ versus the instantaneous kink density $n(t)$, as obtained from numerical calculations (data points); the red line is a guide to the eye and indicates the expected linear dependence. Insets: data for the time evolution of $A(t)$ used in the main plot.
The numerical calculations have been done for a chain with lattice spacing $a=1$, post-quench potential $V(\phi)=0.1(\phi^2-1)^2$, pre-quench thermal ensemble with potential $V_0(\phi) = m_0^2\phi^2/2$, mass $m_0=1$, and initial temperature $\beta_0=18.2$, leading to a final temperature $\beta=8$.
The chain length $L$ is chosen here as indicated in Fig. \ref{fig_5}. 
}
\end{figure}

%
Let us now comment on the expected behavior of the various quantities introduced so far in the phenomenological Eqs.~\eqref{eq_correlation_time}, \eqref{eq_onept_nt_lin}, \eqref{eq_kin}, and \eqref{eq_kin_sol}.
\begin{enumerate}[label=(\roman*)]
\item $A(t)$ introduced in Eq.~\eqref{eq_correlation_time}:
this quantity is the dynamical counterpart of the parameter $A_\beta$ appearing in Eq.~\eqref{eq_twopt_thcor} and it is such that it approaches $A_\beta$ upon increasing $t$.
As we have already discussed in Sec.~\ref{sec_thermo} (see also App. \ref{app_TM}), 
$A_\beta\simeq \bar{\phi}^2$ at low temperatures, while corrections to this equality emerge at finite temperature, due to the presence of radiation (see also Fig.~\ref{fig_2} $(c)$).
Similarly, at short times, the evolution of $A(t)$ is affected by the equilibration of the radiation and it is therefore 
expected to display a non-trivial and non-universal behavior, 
However, after the radiation has relaxed (and therefore Eq.~\eqref{eq_onept_nt} starts to apply), $A(t)$ is completely determined by the time-dependent kink density $n(t)$, as in Eq.~\eqref{eq_onept_nt_lin}, namely $A(t)=A_\beta+\alpha_\beta^{A_\beta}(n(t)-n_\beta)$. 
This relationship is physically motivated by the fact that, as kinks disappear, their energy is converted into radiation, the temperature of which changes in order to ``accommodate'' the new energy. 
As we saw in Section \ref{sec_thermo}, at equilibrium and the in absence of radiation $A=\bar{\phi}^2$, while the presence of radiation renormalizes $A$ to a generically different value close to $\bar{\phi}^2$. In the non-equilibrium case and at long times, one has the same interpretation, but with the amount of radiation changing as kink disappears. As such, $A(t)$ becomes a time-dependent quantity with a linear dependence on the kink's density.
As a numerical check, in Fig.~\ref{fig_A_tampl} we provide a parametric plot of $A(t)$ as a function of the kink density $n(t)$ for various initial conditions, as we did in Fig.~\ref{fig_7} for $\phi^2$ and $\phi^4$. A linear dependence of $A(t)$ on $n(t)$ clearly emerges at sufficiently long times, as noted before for other one-time quantities. Note that this conclusions applies also to other choices of the initial parameters.

\item $\alpha_\beta^O$ introduced in Eq.~\eqref{eq_onept_nt_lin}.
This quantity can be understood as the variation of the expectation value of the local observable $O$ caused by the transformation of a single (anti)kink into radiation (see Fig.~\ref{fig_6}), independently of the presence of other kinks in the state. 
Accordingly, $\alpha_\beta^O$ is mostly determined by the radiation with a resulting algebraic dependence on the temperature, reaching a finite non-zero value for $\beta\to\infty$. 
In the quantum case, the radiation 
has an exponential dependence on the temperature as it decreases and this is expected to 
carry over to $\alpha_\beta^O$.
However, we emphasize that the conclusions we draw here for classical systems carry over to quantum systems only if
the mass of the kink is much larger than that of the radiation.
Accordingly,
$\alpha_\beta^O$ still varies much slower than $n_\beta$ in the quantum case as well.
\begin{figure}[t!]
\includegraphics[width=0.7\columnwidth]{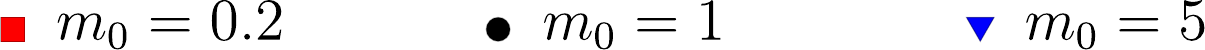}\\ \ \\
\includegraphics[width=0.49\columnwidth]{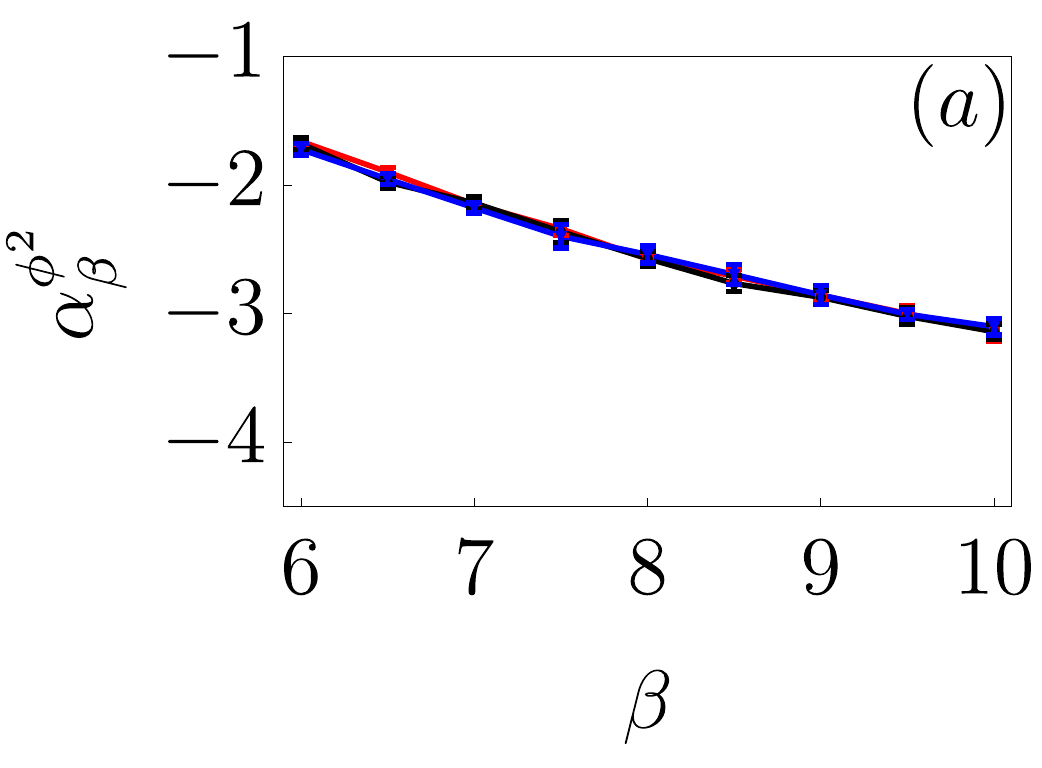}
\includegraphics[width=0.49\columnwidth]{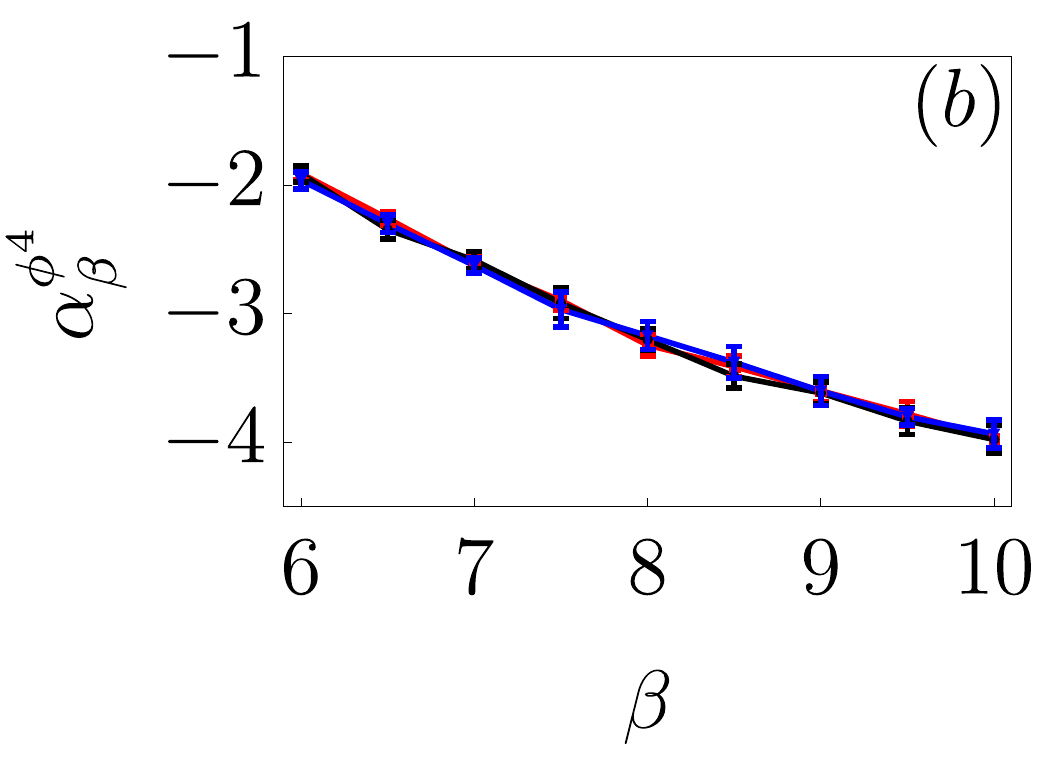}\\ 
\includegraphics[width=0.49\columnwidth]{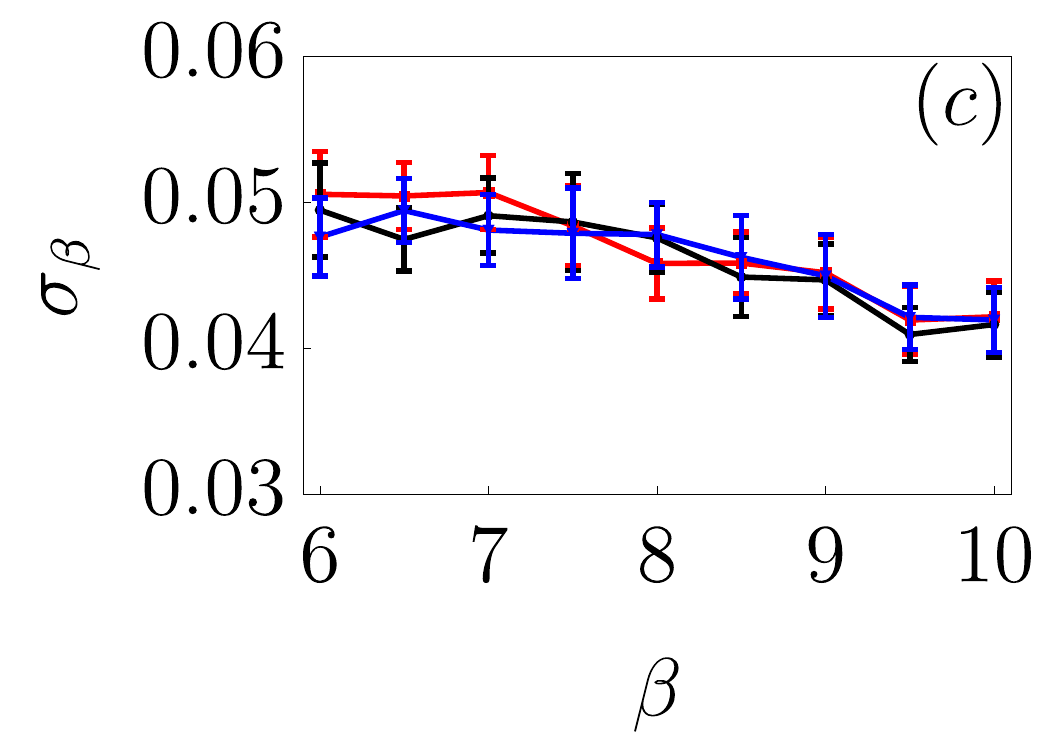}
\includegraphics[width=0.49\columnwidth]{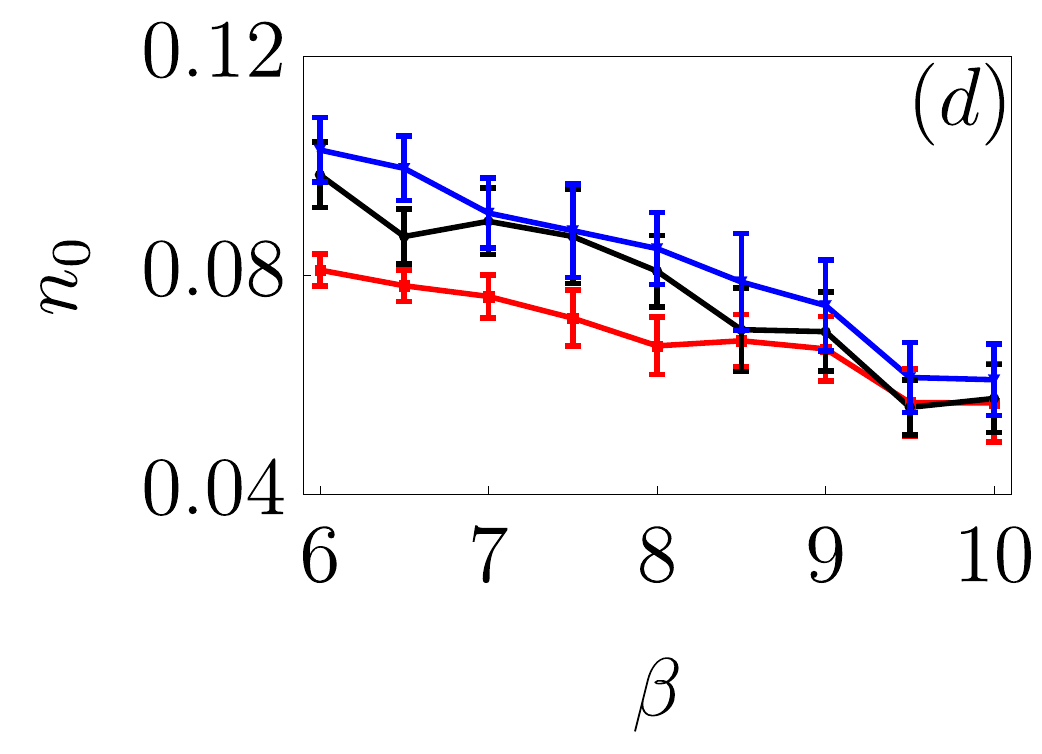}
\caption{\label{fig_9}  (Color online) Numerical values of the constants (a) $\alpha_\beta^{\phi^2}$, (b) $\alpha_\beta^{\phi^4}$, (c) $\sigma_\beta$, and (d) $n_0$ appearing in the phenomenological Eqs.~\eqref{eq_onept_nt_lin} and \eqref{eq_kin}, as functions of the final inverse temperature $\beta$ and for various initial conditions. 
The different colors and symbols correspond to the various values of the pre-quench mass $m_0$, while the initial inverse temperature $\beta_0$ is changed accordingly, in order to result into the inverse temperature $\beta$ indicated in the plot for the final state.
The data shown in the panels, have been obtained with $a=1$ and the post-quench potential $V(\phi)=0.1(\phi^2-1)^2$. 
The fits used to determine numerically (a) $\alpha_\beta^{\phi^2}$, (b) $\alpha_\beta^{\phi^4}$, and (c) $\sigma_\beta$ are done with the least-square method, and the corresponding confidence intervals are determined by requiring that the relative fluctuations of the chi-square are less than $50\%$ 
within the intervals.
}
\end{figure}

In panels (a) and (b) of Fig.~\ref{fig_9} we show $\alpha_\beta^O$, for $O=\phi^2 $ and $O=\phi^4$, respectively, as functions of the final inverse temperature $\beta$, for various initial conditions. 
The numerical values of $\alpha_\beta^O$ are extracted from the late-time behavior of the parametric plot of $O(t)$ as a function of $n(t)$, corresponding to the linear behavior for small $n(t) - n_\beta$ 
(see Fig.~\ref{fig_7}).
The curves corresponding to the different values of $m_0$ are essentially independent of the details of the initial conditions (primarily determined by the initial mass $m_0$, which determines together with $\beta$ also the initial value $\beta_0$ of the inverse temperature) while they depend markedly on the final steady state, as we anticipated above.

\item $\sigma_\beta$ introduced in Eq.~\eqref{eq_kin}.
This cross section can be understood as the rate with which a pair of kinks (the velocity of which is distributed according to a thermal distribution) annihilate into radiation: its dependence as a function of $\beta$ is essentially determined by the corresponding change in the radiative background in which these kinks move, while $\sigma_\beta$ does not significantly depend on their density $n(t)$.
Accordingly, in the classical case, $\sigma_\beta$ is expected to depend algebraically on $\beta^{-1}$ as  $\beta^{-1} \to 0$. $\sigma_\beta$ depends also on the typical kink velocity which, in turn, introduces an additional dependence on $\beta$. 
In the quantum case, instead, the dependence of $\sigma_\beta$ on $\beta^{-1}$ is modified analogously to what happens for the parameter $\alpha_\beta^O$ discussed in the previous point.
In panel (c) of Fig.~\ref{fig_9} we show $\sigma_\beta$ as a function of the inverse temperature $\beta$ for various initial conditions. The corresponding curves exhibit essentially no dependence on the details of the initial conditions, but they depend only on the inverse temperature of the final state.

\item $n_0$ introduced in Eq.~\eqref{eq_kin_sol}.
In panel (d) of Fig.~\ref{fig_9} we show the behavior of $n_0$ as a function of $\beta$ for various initial conditions: 
differently from the quantities $\alpha_\beta^O$ and $\sigma_\beta$ discussed above, 
$n_0$ is not only determined by the final equilibrium state but, as anticipated, it turns out to depend significantly on the initial conditions, each corresponding to a different curve in the plot.
Actually, $n_0$ is the only information on the initial state which continues to be relevant for the long-time dynamics. 
\end{enumerate}

As a final remark, we emphasize that, since the coefficients $\alpha_\beta^O$ and $\sigma_\beta$ turn out to be essentially fixed by the final thermal state, they can be determined in the equilibrium theory and then used to describe the non-equilibrium evolution as  discussed here. 

In practice we recall that $n_\beta$ can be efficiently computed on thermal states by means of the transfer matrix approach, while the computation of $\sigma_\beta$ is a highly non-trivial task, related to the problem of \emph{nucleation}, i.e., on the formation within the system of spatially extended ``bubbles'' corresponding to one of the two vacua.
In fact, based on Eq.~\eqref{eq_kin} and its physical interpretation, one easily realizes that the rate $\Gamma_\beta$ at which kinks are produced in equilibrium by the background fluctuating radiation, i.e., the nucleation rate, is given by $\Gamma_\beta\equiv \sigma_\beta n_\beta^2$.
This problem has been addressed in detail\cite{PhysRevLett.63.2337,PhysRevLett.84.1070,Cattuto_2003,PhysRevB.34.6536,PhysRevLett.60.2563,PhysRevLett.76.2609,PhysRevB.57.7930}
primarily for dissipative systems, in which the field is coupled with some external thermal bath: to the best of our knowledge, there are no analytical results for the nucleation rate in the case of a closed system.

\subsubsection{Comparison with the case of dissipative dynamics}

Several studies have been dedicated to models supporting kink-like excitations but in the presence of a dissipative bath\cite{PhysRevLett.63.2337,PhysRevLett.84.1070,Cattuto_2003,PhysRevB.34.6536,PhysRevLett.60.2563,PhysRevLett.76.2609,PhysRevB.57.7930}. In particular, Ref.~\ocite{PhysRevLett.84.1070} focussed on the dynamics of kinks and their annihilation processes, deriving a kinetic equation for the density of kinks which is close in spirit to our Eq.~\eqref{eq_kin} but differs from it in two respects. 
First, in the presence of dissipation kink and antikink always annihilate when they meet, in clear contrast with the case of isolated system considered here, in which most of the kinks survive the scattering processes (see Fig.~\ref{fig_4}). 
Second, spatial diffusion induces correlations in the annihilation processes. In fact, at low temperatures, the kinks in a kink-antikink pair created by a nucleation process --- which diffuse in space --- are more prone to interact among themselves rather than with a kink belonging to another pair, as the latter would require a significantly longer time compared to the case in which these kinks separate ballistically.
As a consequence, in the presence of dissipation, the annihilation processes do not actually occur among uncorrelated kinks, as the $\propto n^2(t)$ term in Eq.~\eqref{eq_kin} suggests.

However, one generically expects dissipation and diffusion to emerge locally also in an isolated system and therefore one may wonder if and how this affects the dynamics of the kinks we considered.
As a matter of fact, kinks are subject to a random force caused by the fluctuating radiation and diffusion processes effectively take place (see, for instance, the kink trajectory close to the region 4 in Fig.~\ref{fig_4}, lower panel).
However, from our numerical study it turned out that the dissipation is not strong enough to cause diffusive behavior on a lengthscale comparable with the average kink distance, thus ensuring the validity of the kinetic equation Eq.~\eqref{eq_kin} in the regime that we focused on.

The reason for this can be better understood as follows. 
The diffusive behavior of a kink in between two scattering events is determined by the amount of radiation,
which becomes algebraically small upon decreasing the temperature in a classical system (cf.~Eq.~\eqref{eq_chi_corr_quantum} and discussion thereunder), the diffusion length-scale $\ell_D$
is also expected to decrease algebraically upon decreasing the temperature. 
On the other hand, the kink density is exponentially suppressed upon decreasing the temperature, i.e., upon increasing $\beta^{-1}$ and consequently the average distance $n_\beta^{-1}$ between consecutive (anti)kinks grows exponentially with it. 
Accordingly, for temperatures small enough, $\ell_D$ becomes shorter than the kink average distance,
much smaller than $n_\beta^{-1}$ and the kink motion becomes diffusive.   
However, this condition requires an extremely low kink' density 
which (i) is negligible for the dynamics of one-point observables and (ii) is not accessible within the system sizes $L$ which we can simulate, violating the condition  $L n(t)\gg 1$ which has to be satisfied in order to attain the thermodynamic limit (see the discussion in Sec.~\ref{subsec_quench} and in App. \ref{app_Met}).
As a result, diffusive corrections are negligible within the range of values of parameters and timescales we investigated in the present study. Indeed, the predictions of the kinetic equation \eqref{eq_kin} fit remarkably well our numerical data (see Fig. \ref{fig_8}).

\subsubsection{Interrupted aging}

As anticipated in Sec.~\ref{sec_kineq}, 
the behavior of the order parameter correlation $\langle \phi(t,x)\phi(t,y)\rangle$ at times $t\ll (2n_\beta \sigma_\beta)^{-1}$ exhibits the scaling form (cf.~Eqs.~\eqref{eq_correlation_time} and \eqref{eq_power_law})
\be\label{eq_coarsening}
\langle \phi(t,x)\phi(y,t)\rangle \sim f\left(\frac{|x-y|}{\sigma_\beta t}\right),
\ee
with $f(x) \propto \exp(-2x)$. This is the same scaling form as the one predicted for the \emph{coarsening dynamics} of 
systems quenched across the critical 
temperature\cite{Bray1994,Godreche2002,Cugliandolo2015}. 
In these systems, although the global symmetry cannot be dynamically broken by the quench, this occurs\emph{locally} via the creation of domains within which the order parameter $\phi$ takes the value corresponding to one of the possible vacua.
The average linear size $n^{-1}(t)$ of the ordered domains 
increases algebraically upon increasing $t$ as $n^{-1}(t) \propto t^{1/z_\text{d}}$, where $z_\text{d}>0$ is an exponent which depends on the universal properties of the model: equilibrium is eventually reached in a time scale increasing upon increasing the linear system size $L$. 
This lack of a length scale of microscopic origin, i.e., intrinsic to the system, affects the equal-time two-point correlation function of the order parameter, giving rise to the scaling form~\eqref{eq_coarsening}. 

However, the scaling behavior in Eq.~\eqref{eq_coarsening} emerges only if the eventual steady state can host a stable ordered phase, which is not the case of the system considered in this work because of its reduced spatial dimensionality. 
In fact, the model considered here is expected to relax to an equilibrium state at finite temperature, which cannot sustain order in one spatial dimension. 
At times $t \gtrsim (2n_\beta \sigma_\beta)^{-1}$, in fact, the scaling in Eq.~\eqref{eq_coarsening} is no longer valid, as the domain size $ n(t)^{-1}$ saturates to a finite value. This is analogous to dynamics after a quench right above the critical temperature, in a system sustaining a phase transition: in that case, domain's size is expected to grow until it reaches the equilibrium correlation length $\xi_{eq}$. In this sense, our model undergoes an interrupted aging dynamics.

\section{Conclusions and outlook}
\label{sec_concl}

In this work we investigated the late-time dynamics of one-dimensional closed systems featuring topological excitations. Our findings show that the non-equilibrium large distance properties of these system display universality, in that they can be described solely in terms of a simple effective dynamics of the kinks. So far, this was only known for the low-temperature thermodynamics of these models, while we demonstrated that these features persist also in non-equlibrium protocols.

Initializing the system in an non-equilibrium $\mathbb{Z}_2-$even ensemble, resulting in an initial large number of kinks, universal features emerge at late times when the kink density decreases.
We found that the time-evolving density of kinks determines (i) the correlation length of the two-point (equal time) correlator of the order parameter and (ii) the approach to equilibrium of one-point $\mathbb{Z}_2-$even observables, which display a linear dependence on the instantaneous kink density.

The roots of such a behavior can be tracked down in a separation of time scales between kinks and radiative modes: the kink density relaxes much slower to its thermal value compared with the radiation, which then acts as a thermal bath on these excitations. As such, the total kink density emerges as the only degree of freedom controlling the late-time behavior.
A simple kinetic equation describes the late-time evolution of the kinks' density, accounting for the possibility of annihilating pairs of kinks 
and, conversely, their creation from the background fluctuating radiation.

Several interesting directions are left for future investigations. 
First, while analytical predictions exist for the thermal kink density, we are not aware of results concerning the thermal nucleation rate, which is ultimately related to the cross section appearing in the kinetic equation, and which we treated as a fitting parameter. Being able to determine such a cross-section would further boost the predictive power of our phenomenological model.
Another interesting direction concerns the possibility of refining our approach, by devising a whole phenomenological model able to capture the entire time evolution of the kinks. Indeed, within this work, we focused on the total kinks' density resulting in a mean field description. On the other hand, controlling the details of the dynamics of the kinks (e.g., their trajectories) is necessary, for example, in determining two-time correlation functions of the order parameter (see Refs.~\ocite{PhysRevE.93.062101,PhysRevLett.119.100603,PhysRevB.100.035108,PhysRevLett.119.010601} for the integrable case).
Finally, the study of weak confinement of the topological excitations is a compelling quest, with possible connections with the physics of quantum scars\cite{Turner2018,PhysRevB.99.161101,PhysRevLett.122.040603,PhysRevLett.122.173401,PhysRevB.98.155134,PhysRevLett.122.220603}.

\subsection*{Acknowledgements}
We thank S.~Diehl, M.~Kormos, A. Langins, and N.~Robinson for very useful discussions. 
AB acknowledges support from the European Research Council (ERC) under ERC Advanced grant 743032 DYNAMINT. AC acknowledges support from the ERC under the Horizon 2020 research and innovation programme, grant agreement No. 647434 (DOQS). This research was supported in part by the National Science Foundation under Grant No. NSF PHY-1748958. 

\appendix

\section{The transfer matrix approach to thermal expectation values}
\label{app_TM}

\begin{figure*}[t!]
\includegraphics[width=0.3\textwidth]{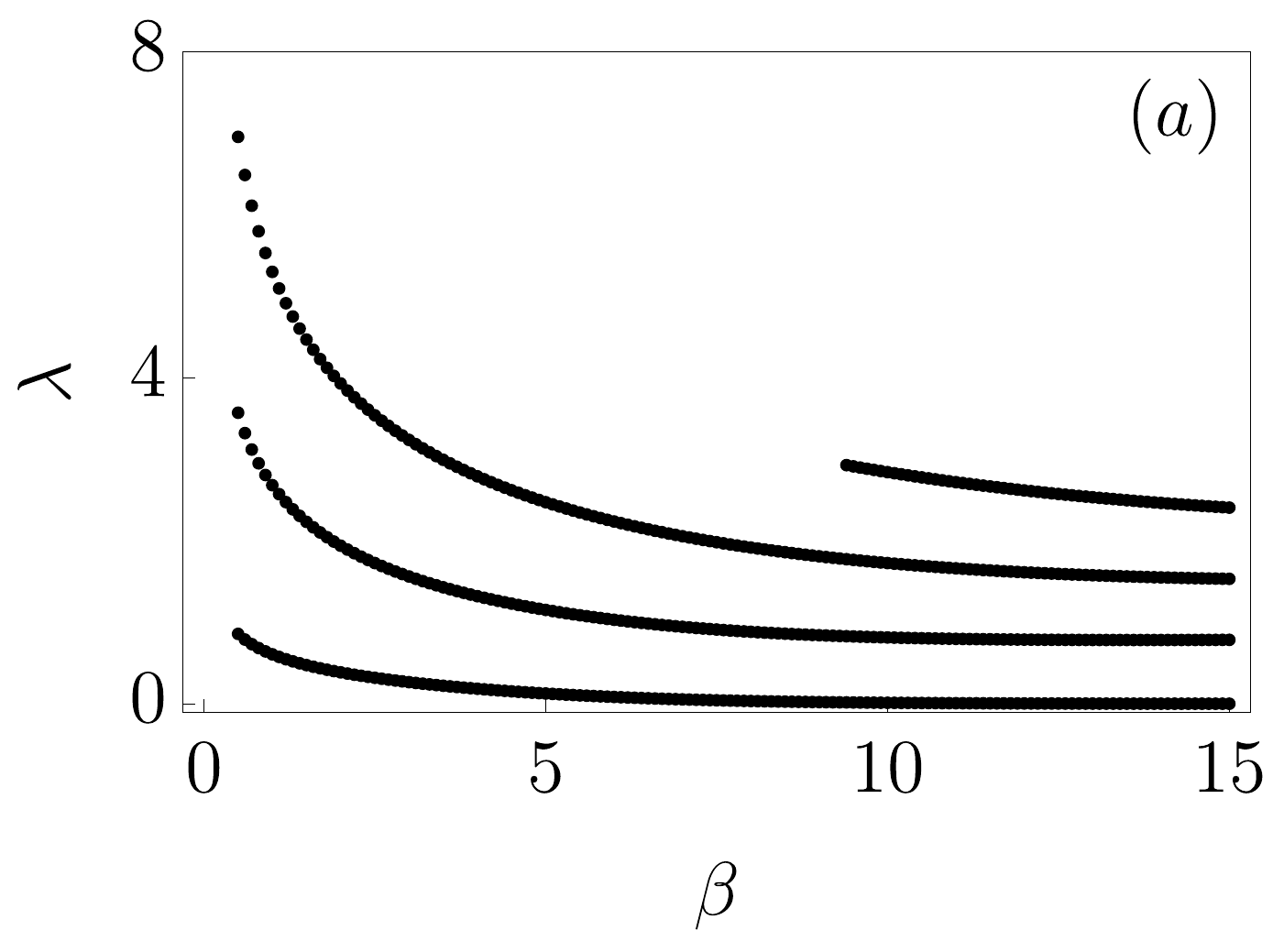}\,\,
\includegraphics[width=0.3\textwidth]{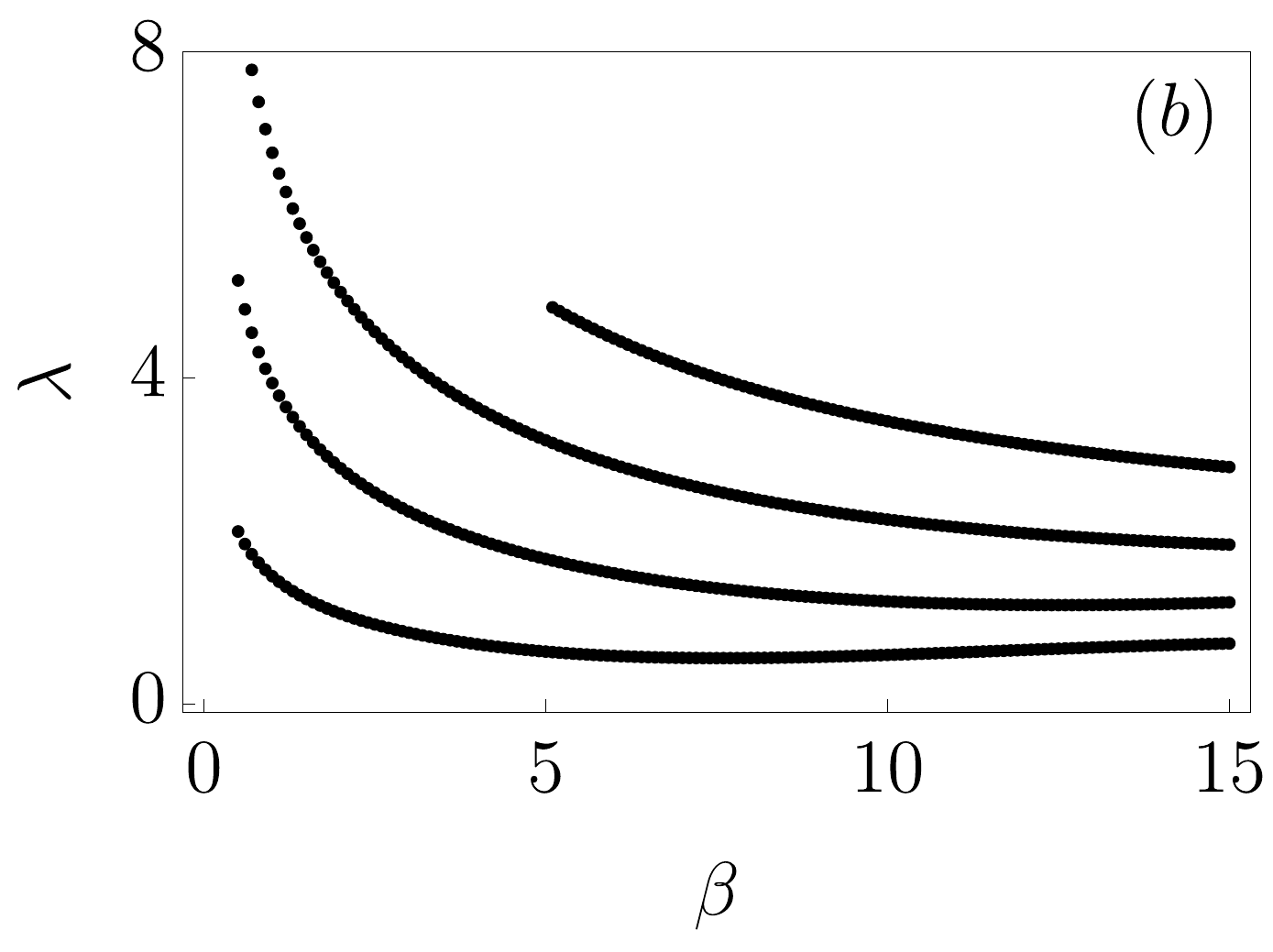}\,\,
\includegraphics[width=0.31\textwidth]{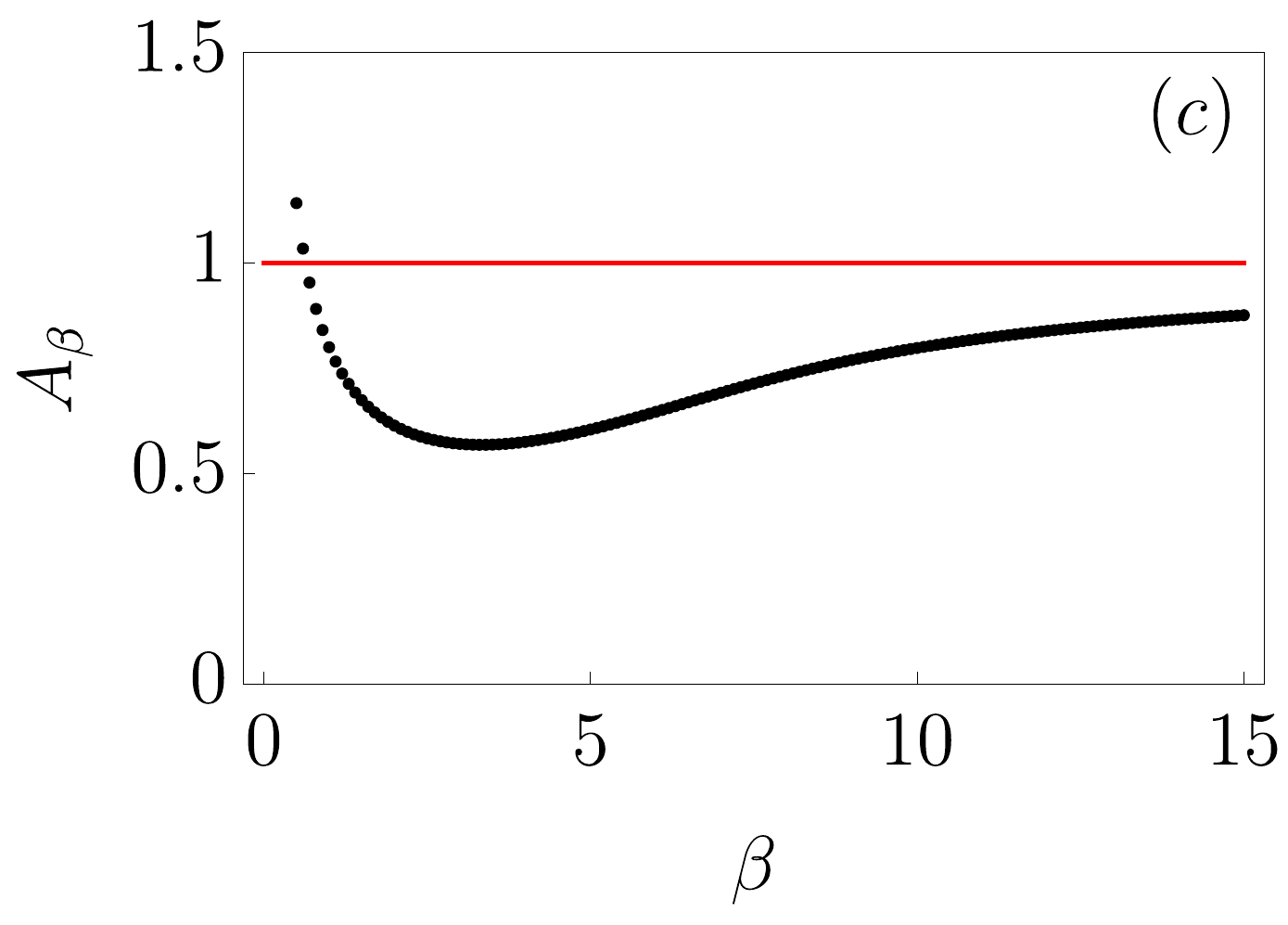}\,\,
\caption{\label{fig_2}(Color online) Transfer matrix results for correlation functions at finite temperature for the continuum model 
defined in Eq.~\eqref{eq_H_continuum}. 
Panel $(a)$ (resp. $(b)$): correlation lengths $\{\lambda_i\}$ in the Laplace transform Eq. \eqref{eq_Laplace} for the most relevant Laplace coefficients $(|L_O(\lambda_i)|>10^{-5})$ for the correlation $\langle\phi(x)\phi(y)\rangle_c$ (resp. $\langle\phi^2(x)\phi^2(y)\rangle_c$) as a function of the inverse temperature $\beta$. 
In the low-temperature limit, in panel $(a)$ one curve clearly approaches zero, resulting in a divergent correlation length for $\langle\phi(x)\phi(y)\rangle_c$ according to Eq.~\eqref{eq_Laplace}. This feature is absent in $\langle\phi^2(x)\phi^2(y)\rangle_c$ (see panel $(b)$).
Panel $(c)$: value of the Laplace coefficient associated with the largest correlation length in $\langle\phi(x)\phi(y)\rangle_c$, which is interpreted as the prefactor $A_\beta$ in Eq.~\eqref{eq_twopt_thcor}, as a function of the inverse temperature $\beta$. All the results are obtained with $V(\phi)=0.1(\phi^2-1)^2$ in the continuum limit Eq. \eqref{eq_H_continuum}. Indeed, in the low-temperature limit $A_\beta$ approaches $\bar{\phi}^2$ (red line), which for this potential is $\bar{\phi}=1$.
}
\end{figure*}

The transfer matrix approach allows a straightforward computation of observables on classical thermal states, taking advantage of the correspondence between a one-dimensional classical problem and a quantum one. Within this Appendix, we summarize the ideas presented in Ref. \ocite{PhysRevB.6.3409,PhysRevB.11.3535,PhysRevB.22.477,PhysRevE.48.4284}.
Let us consider expectation values on a thermal ensemble based on the Hamiltonian~\eqref{eq_H}: a finite chain of $N$ sites with periodic boundary conditions. We will consider the thermodynamic limit $N\to \infty$. On thermal ensembles, the momentum $\Pi(x)$ is completely uncorrelated with the field $\phi(x)$.
For the sake of simplicity, let us focus on observables that depend only on the field $\phi$.
Their correlation functions can be written as
\begin{multline}
\langle \prod_{j=1}^s O_j(\phi(x_j))\rangle= \frac{1}{\mathcal{Z}[\beta]}\int \dd ^N \phi \prod_{j=1}^s O_j(\phi(x_j))\times \\ e^{-\beta a\sum_{x=1}^N\Big[\frac{1}{2 a^2}(\phi(x+a)-\phi(x))^2+V(\phi(x))\Big] }
\, .
\end{multline}
We account for possibly different observables introducing a label.
We now read the above equation
as a trace over a product of suitable operators on an Hilbert space. We define ket (and bra) states $\ket{\phi}$ with the normalization condition $\langle\phi'|\phi\rangle=\delta(\phi-\phi')$. We promote the observables from classical objects to operators $O_j\to \hat{O}_j$ with diagonal entries
\be
\langle\phi'|\hat{O}_j|\phi\rangle=\delta(\phi-\phi')O_j(\phi)
\; .
\ee
Furthermore, we define a transfer matrix $\hat{\tau}$ with the following elements
\be
\!\!\! \langle\phi'|\hat{\tau}|\phi\rangle = e^{-\frac{\beta a}{2} \left[a^{-2}(\phi-\phi')^2+V(\phi)+V(\phi')\right]}
\; .
\ee
In this language, the partition function simply becomes $\mathcal{Z}[\beta]=\text{Tr} \hat{\tau}^N$, the expectation value 
of a single local observable reads
\be
\langle O(\phi(x))\rangle=\frac{\text{Tr}[\hat{\tau}^N\hat{O}]}{\text{Tr}[\hat{\tau}^N]}
\; , 
\ee
 two-point correlation functions are
\be
\langle O_1(\phi(x)) O_2(\phi(x+y))\rangle=\frac{\text{Tr}[\hat{\tau}^{N-y}\hat{O}_1\hat{\tau}^y \hat{O}_2]}{\text{Tr}[\hat{\tau}^N]}
\; ,
\ee
and so on and so forth.
$\hat{\tau}$ is clearly a Hermitian operator, which can be diagonalized and has real spectrum. Let $\ket{\ell }$ be its orthonormal eigenvectors with corresponding eigenvalue $\mu_\ell$, i.e., $\hat{\tau}|\ell\rangle=\mu_\ell |\ell\rangle$. The operator $\hat{\tau}$ is bounded both from below and above (we assume $V(\phi)\ge 0$) $0\le \langle \phi'|\hat{\tau}|\phi\rangle\le 1$, thus $0<\mu_\ell\le 1$. We order $\mu_\ell>\mu_{\ell+1}$ and call the eigenstate $\ell=0$ the ground state, which we assume to be unique (and later in the continuum limit justify this claim).
In the thermodynamic limit $N\to\infty$, the ground state is the state contributing the most to the partition function $\mathcal{Z}[\beta]=\mu_0^N\left[1+\mathcal{O}((\mu_1/\mu_0)^N)\right]$. Accordingly, in this limit we can compute correlation functions by 
simply projecting on the ground state
\be
\langle O_1(\phi(x)) O_2(\phi(x+y))\rangle=\frac{\langle 0|\hat{O}_1\hat{\tau}^y \hat{O}_2|0\rangle}{\mu_0^y}\, .
\ee
Inserting now a representation of $\hat{\tau}$ in the diagonal basis, we have
\begin{eqnarray}
\label{eq_lap_tr_noc}
&&
\langle O_1(\phi(x)) O_2(\phi(x+y))\rangle
\nonumber\\
&& \quad\quad
=
\sum_{\ell=0}^\infty \langle 0|\hat{O}_1|\ell\rangle\langle \ell| \hat{O}_2|0\rangle \left(\frac{\mu_\ell}{\mu_0}\right)^y\, .
\end{eqnarray}
In particular, the connected correlation function is readily recovered by subtracting the $y\to\infty$ limit.
\begin{eqnarray}
\label{eq_lap_tr}
&& 
\langle O_1(\phi(x)) O_2(\phi(x+y))\rangle_c 
\nonumber\\
&& 
\quad\quad
= \sum_{\ell=1}^\infty \langle 0|\hat{O}_1|\ell\rangle\langle \ell| \hat{O}_2|0\rangle e^{-y\log\left(\frac{\mu_0}{\mu_\ell}\right)}\, .
\end{eqnarray}
Accordingly, the diagonalization of the transfer matrix $\hat{\tau}$ gives immediate access to one-point expectation values and to the Laplace transform
\be\label{eq_Laplace}
\langle O(x)O(y)\rangle_c =\sum_{i=1}^\infty e^{-\lambda_i |x-y|}L_O(\lambda_i)
\ee
 of the correlation functions. 

From Eq.~\eqref{eq_lap_tr} we can connect the divergence of a correlation length with the closure of the gap between the ground state and the first excited state $\mu_1\to\mu_0$. This can be most easily understood in the continuum limit. Taking $a\to 0$, we approximately get
\be
\langle\phi'|\hat{\tau}|\phi\rangle\simeq \langle\phi'|\left(1-a\hat{\mathcal{H}}_q\right)|\phi\rangle\simeq \langle\phi'|e^{-a \hat{\mathcal{H}}_q}|\phi\rangle\, ,
\ee
with $\hat{\mathcal{H}}_q$ the zero-dimensional Hamiltonian
\be\label{eq_H_quan}
\hat{\mathcal{H}}_q=\beta\frac{\hat{p}^2}{2}+\beta V(\hat{\phi})\, ,
\ee
where we introduced the canonically conjugate operators $[\hat{\phi},\hat{p}]=i$.
The spectrum of $\hat{\mathcal{H}}_q$ gives the correlation lengths. 
In particular, we are interested in the energy gap between the ground state and the first excited state. In the Hamiltonian~\eqref{eq_H_quan} it is convenient to perform a canonical transformation $\hat{p}\to \beta^{-1/2}\hat{p}$ and 
$\hat{\phi}\to \beta^{1/2} \hat{\phi}$, which leads to the new Hamiltonian
\be
\hat{\mathcal{H}}_q\to \hat{\mathcal{H}}_q=\frac{\hat{p}^2}{2}+\beta V(\beta^{-1/2}\hat{\phi})
\, .
\ee

Let us now consider, for example, the double-well potential $V(\phi)=\lambda (\phi^2-\bar{\phi}^2)^2$. The potential appearing in the quantum mechanical effective problem has the same qualitative form, with a potential barrier increasing as $\beta\to \infty$.

Firstly, since the potential is unbounded from above we get that the energy levels are discrete (i.e., the Laplace transform~\eqref{eq_lap_tr_noc} is indeed a discrete summation) and non degenerate. Then, the eigenvectors are alternatively even and odd with 
respect to the $\mathbb{Z}_2$ symmetry: in particular, the ground state is even and the first excited state is odd.
The parity of the ground state guarantees that the one-point function of odd powers of the field vanishes, as it should be.

Let us now increase $\beta$: the potential barrier progressively grows and the first excited state becomes degenerate with the ground state. Instead, a finite gap remains between the ground state and the second excited state.
The closure of the first gap, but not of the second, implies that there is one and only one divergent correlation length in Eq.~\eqref{eq_lap_tr}. Such a correlation length can be associated with the inverse of the kink density.

Indeed, this has been used to analytically study the low-temperature limit of the density of kinks\cite{PhysRevB.6.3409,PhysRevB.11.3535,PhysRevB.22.477,PhysRevE.48.4284}.

\begin{figure}[b!]
\includegraphics[width=0.9\columnwidth]{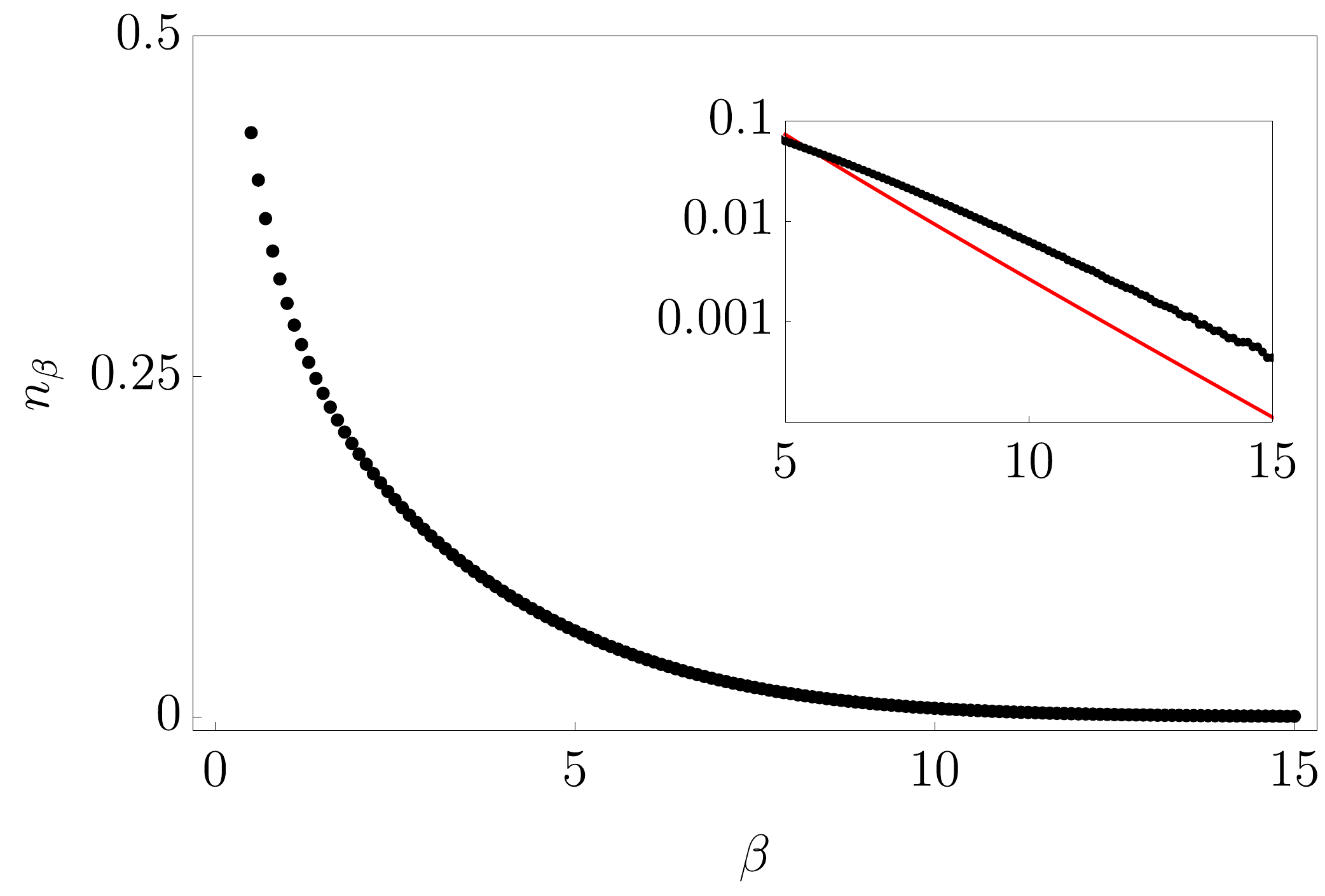}
\caption{\label{fig_2_bis} (Color online) Thermal kink density as a function of the inverse temperature $\beta$, extracted from the numerical evaluation of the correlator order parameter using Eq.~\eqref{eq_twopt_thcor}. The largest correlation length is numerically computed through transfer matrix methods (App. \ref{app_TM}).  Inset: comparison between the numerical result (black line) and the asymptotic expansion Eq.~\eqref{eq_n_beta} (red line), vertical axis in logarithmic scale. 
}
\end{figure}
In Fig.~\ref{fig_2} the most relevant Laplace coefficients $|L_O(\lambda_i)|\ge 10^{-5}$ are reported for the correlation functions of $\phi(x)$  [panel (a)] and its square $\phi^2(x)$ [panel (b)] in the continuum limit $a\to 0$.
For $\phi(x)$, one curve clearly approaches zero upon increasing $\beta$, resulting in a divergent correlation length which can be associated with the density of kinks as dictated by Eq.~\eqref{eq_twopt_thcor}.
For $\phi^2$, instead, the curves do not approach zero: the correlation function is not sensible to the presence of kinks. In a first approximation, in this case the correlation is due to the radiative modes and thus the correlation length remains finite as the temperature is decreased.
In panel (c) we show the extrapolated value of $A_\beta$ as defined in Eq.~\eqref{eq_twopt_thcor}. As $\beta$ is increased, $A_\beta$ slowly approaches the value determined by the bare vacuum (the red line corresponding to $1$). Departures from this value are interpreted as renormalization effects due to the radiative modes.

In Fig.~\ref{fig_2_bis} we show the kink density extracted from the spatial correlation of the order parameter as a function of the inverse temperature $\beta$. We observe a good exponential decay qualitatively in agreement with Eq.~\eqref{eq_n_beta}, with a finite-temperature renormalization of the kink mass, changing the characteristic scale of the exponential decay.
We conclude this Appendix by noticing that, for parity reasons, the only non-vanishing matrix element in the form $\langle 0|\hat{\phi}^n|1\rangle$  occur for odd power  $n$. Indeed, only correlators of odd powers of the field are sensitive to the kink density, which causes the correlation length to diverge.

\section{Real-time numerical simulation}
\label{app_Met}

Here, we describe the numerical method used for a first-principle exploration of the dynamics. 

The algorithm consists of two steps: first we randomly generate a field configuration from the desired ensemble, then the initial condition is deterministically evolved in time and the desired observables are computed along the time evolution. The operation is repeated and observables are averaged over the initial conditions (we took advantage of translational invariance and averages also on lattice sites). 

We consider a system of $L$ sites with periodic boundary conditions, $L$ must be taken large in order to ensure the thermodynamic limit. We note that we can approximately realize the thermodynamic limit only if a large number of kinks is present in the system: as a rule of thumb we ask $L n(t)\gtrsim 10$, with $n(t)$ the instantaneous average kink density. 
In order to be able to study the whole relaxation process until the thermal ensemble is reached, we therefore need $L \gtrsim 10 n_\beta^{-1}$. Since $n_\beta^{-1}$ is exponentially divergent upon decreasing the temperature, this limit can be hard to achieve. Thus, for very low temperatures we did not follow the whole thermalization process. Instead, fixing a large $L$ (we consider up to $L=2^{11}$ points), we stopped the time evolution when $n(t)$ became too small. 

The average number of kinks present in the system also affects the required number of initial configurations to be sampled: we found that, by choosing the system length $L$ in such a way that $L n(t)\gtrsim 10$, the average on c.a. $4000$ independent realizations is a good compromise between speed and accuracy of the numerical data.
Any given initial field configuration $\phi(t,x)$ is evolved according to the second-order (in the time step discretization) symplectic algorithm used in Ref.~\ocite{De_Luca_2016}
\begin{eqnarray}
&& \phi(t+\dd t,x) 
= 
\frac{\dd t^2}{a^2} (\phi(t,x+a)+\phi(t,x-a))
\nonumber\\
&&
\qquad 
-\phi(t-\dd t,x)+2 \phi(t,x)\left(1-(\dd t/a)^2\right)
\\
&& 
\qquad
-\dd t^2 V'(\phi(t,x))\, .
\nonumber
\end{eqnarray}
Higher-order algorithms exist, but we experienced that this second-order one is a good compromise between accuracy and speed.
Initial field configurations are sampled from a free thermal ensemble with mass $m_0$: this simple choice allows us to 
quickly generate random initial conditions. Indeed, on a free ensemble the Fourier modes are independent Gaussian variables. 
Accordingly, we assign to the field and its derivative the following initial expressions
\begin{eqnarray}
&& 
\phi(x)= \frac{1}{\sqrt{L a}}\sum_{k=0}^{L-1} \frac{1}{\sqrt{2E(2\pi k/L)}}
\nonumber\\
&& \qquad\qquad\qquad \times \left(e^{i2\pi  k x/L}A(k)+\text{c.c.}\right)\, ,
\end{eqnarray}
\begin{eqnarray}
&& \partial_t\phi(x)= \frac{1}{\sqrt{L a}}\sum_{k=0}^{L-1} \sqrt{\frac{E(2\pi k/L)}{2}}
\nonumber\\
&&
 \qquad\qquad\qquad \times 
\left(-ie^{i 2\pi k x/L}A(k)+\text{c.c.}\right)\, .
\end{eqnarray}
Above, $\text{c.c.}$ stands for the complex conjugate and $E(q)$ is the free dispersion law $E(q)=\sqrt{2a^{-2}(1-\cos q)+m_0^2}$.
The modes $A(k)$ are random Gaussian variables with zero mean and variance
\be
\langle A(k)A^*(q)\rangle=\delta_{k,q}\frac{1}{\beta_0 E(2\pi k/L)}\, .
\ee

\bibliography{biblio}

\end{document}